\documentclass[a4paper,10pt,reqno]{amsart}

\usepackage{a4wide}
\usepackage{hyperref}
\usepackage{amssymb}
\usepackage{eucal}
\usepackage{eufrak}
\usepackage{graphics}
\usepackage{graphicx}
\usepackage{epsfig}
\usepackage{multicol}
\usepackage{xypic}
\usepackage{amsmath}
\usepackage{wasysym}
\usepackage{dsfont}

\usepackage{eufrak}
\usepackage{graphics}
\usepackage{graphicx}
\usepackage{epsfig}
\usepackage{multicol}
\usepackage{xypic}
\usepackage{amsmath}
\usepackage{color}

\newtheorem{definition}{Definition}[section]

\newtheorem{remark}[definition]{Remark}

\newtheorem{theorem}[definition]{Theorem}

\newcommand{\lp}{\left(}
\newcommand{\rp}{\right)}
\newcommand{\lc}{\left\{}
\newcommand{\rc}{\right\}}
\newcommand{\der}{\partial}

\newcommand{\bra}{\langle}
\newcommand{\ket}{\rangle}
\newcommand{\R}{\mathbb{R}}      

\newcommand{\N}{\mathbb{N}}      

\newcommand{\F}{\mathds{F}}

\newcommand{\Flder}{\rightarrow}

\usepackage{amssymb}

\newcommand{\proa}{A^*G \mbox{$\;$}_{\tau^*} \kern-3pt\times_\alpha
G \mbox{$\;$}_\beta \kern-3pt\times_{\tau^*} A^*G}

\begin{document}

\title[Continuous and discrete damping reduction]{Continuous and discrete damping reduction for systems with quadratic interaction}

\author[F. Haddad Farshi]{Farhang Haddad Farshi} 

\author[F. Jim\'enez]{Fernando Jim\'enez} 

\author[S. Ober-Bl\"obaum]{Sina Ober-Bl\"obaum}


\maketitle

\vspace{-0.6cm}

\begin{center}
{\bf Department of Engineering Science, University of Oxford}\\
{\bf Parks Road, Oxford OXI 3PJ, UK}\\
\vspace{0.2cm}
$\,\,\,\,\,\,\,\,\,$ e-mail: farhang.haddadfarshi@eng.ox.ac.uk\\
e-mail: fernando.jimenez@eng.ox.ac.uk\\
$\,\,\,\,$e-mail: sina.ober-blobaum@eng.ox.ac.uk
\end{center}

\begin{abstract}
We study the connection between Lagrangian and Hamiltonian descriptions of closed/open dynamics, for a collection of particles with quadratic interaction (closed system) and a sub-collection of particles with linear damping (open system). We consider  both continuous and discrete versions of mechanics. We define the Damping Reduction as  the mapping from the equations of motion of the closed system to those of the open one. As variational instruments for the obtention of these equations we use the Hamilton's principle (closed dynamics) and Lagrange-d'Alembert principle (open dynamics). We establish the commutativity of the branches {\it Legendre transform + Damping Reduction} and {\it Damping Reduction+Legendre transform}, where the Legendre transform is the usual mapping between Lagrangian and Hamiltonian mechanics. At a discrete level, this commutativity provides interesting insight about the  resulting integrators. More concretely, Discrete Damping Reduction yields particular numerical schemes for linearly damped systems which are not symplectic anymore, but preserve some of the features of their symplectic counterparts from which they proceed (for instance the semi-implicitness in some cases). The theoretical results are illustrated with the examples of the heat bath and transmission lines. In the latter case some simulations are displayed, showing a better performance of the integrators with variational origin.
\end{abstract}

\section{Introduction}

The variational description of non-conservative physical systems (open systems) represents a difficult task from the mathematical physics perspective. A cornerstone concerning this issue is the proof in \cite{Bauer} that the dynamical equations of a linearly damped mechanical system cannot be obtained via the Hamilton's principle \cite{AbMa,Arnold}. Ever since, there have been several attempts to provide a general method of dealing with non-conservative forces in classical mechanics, for instance the use of Rayleigh dissipation forces \cite{Goldstein}, the doubling of variables \cite{ Bateman,Galley,David} and the use of fractional derivatives \cite{Cresson, JiOb, Riewe}. A remarkable approach, due to its phenomenological versatility, is the Lagrange-d'Alembert principle \cite{Bloch}, where the variation of the action is set equal to an additional integral term involving the work done by external forces (those provoking that the system is not conservative) under virtual displacements. As happens with the usual Hamilton's principle (which is suitable for the dynamical description of conservative systems), the Lagrange-d'Alembert principle can be performed  in Lagrangian and Hamiltonian fashions, both related to each other by the Legendre transformation \cite{AbMa}. It is important to note, however, that Lagrange-d'Alembert principle is not variational in the pure sense on the word.

Among all non-conservative systems, we will focus  in this work on those subject to linear damping. From a physical perspective, damping can be considered as a phenomenon produced by the interaction of the open system under consideration and its environment, forming both ({\it open system+environment}) a closed system which can be dealt with by means of the usual variational techniques. This closed-system modeling of open system's dynamics is widely used in describing dissipative quantum mechanical systems \cite{Breuer}. As for the closed system, we consider a collection of particles where we allow quadratic interaction. The open subsystem will be a sub-collection of these particles subject to linear damping (and also to a conservative potential interaction among them). We refer to the process of Damping Reduction (DR)  as the mapping from the equations of motion of the closed system  obtained from the Hamilton's principle (i.e.~Euler-Lagrange equations and Hamilton equations) to those expressing  the external forcing (yield by Lagrange-d'Alembert principle).

\begin{figure}[!htb]\label{Sketch1}
\begin{tabular}{cc}
\includegraphics[scale=0.37]{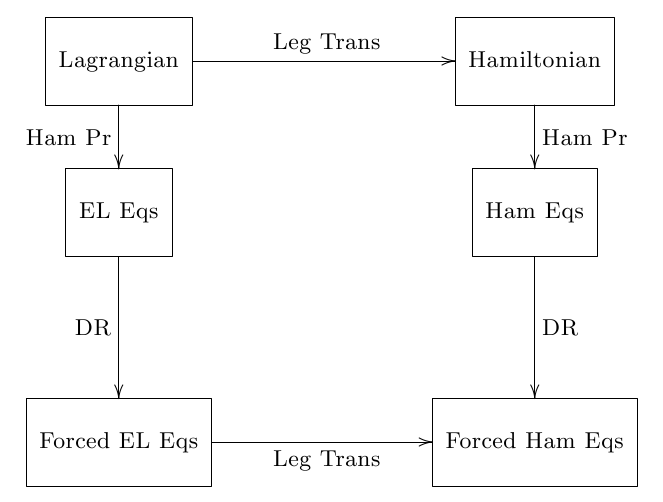}&  \includegraphics[scale=0.37]{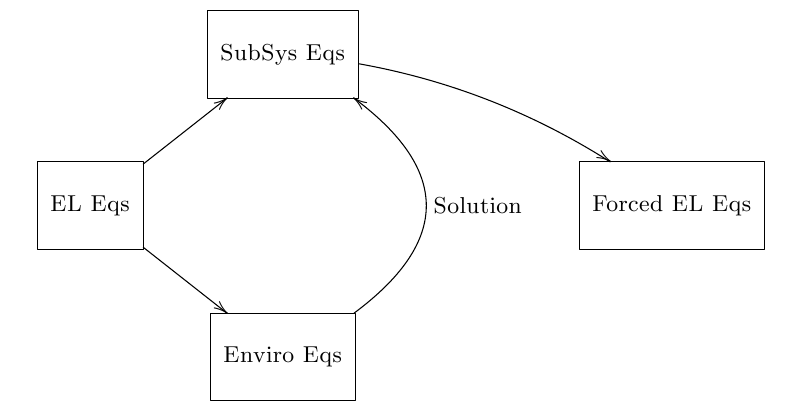}
\end{tabular}
\caption{Left: scheme of the commutativity between the Legendre transformation and the Damping Reduction in Lagrangian and Hamiltonian sides. Right: scheme of the Damping Reduction process at a dynamical level.}
\end{figure}

  The main aim of this paper is showing the commutativity of the  diagram schematically displayed in Figure \ref{Sketch1} (left), which involves Hamilton's principle, the Legendre transformation and DR.  It is worth noting that the path EL Eqs $\Flder$ Forced EL Eqs $\Flder$ Forced Ham Eqs, shall involve as well the Lagrange-d'Alembert principle, as the pseudo-variational tool to describe forced mechanical systems. We remark also that the DR is performed from the dynamical equations, and not at the action level. In particular, we differentiate between the dynamical equations of the open system and the environment; replacing the solution of the latter (which will depend in general on the initial conditions of all variables) in the former, leads to the forced equations of the open subsystem.  This is schematically shown in Figure \ref{Sketch1} (right) for the Lagrangian side (the Hamiltonian is equivalent).
Finally, our investigation not only concerns the continuous version of mechanics, but also the discrete one.

The discrete version of mechanics \cite{MaWe,MoVe} yields unified numerical
schemes approximating the continuous dynamics with powerful structure-preserving properties. In particular, the discretisation of Hamilton's principle provides the so-called discrete Euler-Lagrange equations, which under some regularity conditions produce numerical integrators (variational integrators) that are symplectic \cite{SS} and momentum preserving in the presence of symmetries.  These integrators are applied over conservative systems and, in spite they do not preserve the energy,  have proven to present a stable behaviour,  explained in terms of Backward Error Analysis \cite{BEA,HLW}.  Both the Legendre transformation \cite{MaWe} and the Lagrange-d'Alembert principle  \cite{ObJuMa} are available at the discrete level (represented the former as discrete momentum matching; accounting the latter for the discrete description of externally forced systems, hence open), and therefore the question whether the diagram Figure \ref{Sketch1} (left) is commutative for discrete mechanics raises naturally. We investigate this circumstance for systems with quadractic interaction and a continuous family of discretisations depending on a parameter $\gamma\in[0,1]\subset\R$. We  prove the commutativity; moreover, we observe that the numerical schemes after the discrete version of the Damping Reduction, although non-symplectic anymore (they are applied now to open systems), preserve some of the features of their symplectic couterparts, for instance the semi-implicitness in some cases. Furthermore, we illustrate our results with the example of two particles with quadratic potential coupled to transmission lines, and observe a superior approximation of the dissipative energy in the case of the obtained semi-implicit schemes over other state-of-the-art schemes, namely the implicit and explicit Euler methods.

\vspace{0.4cm}

The paper is organised as follows: \S\ref{CLHDesc} and \S\ref{DiscreteFramework} introduce Hamilton's principle and the Lagrange-d'Alembert principle, both in Lagrangian and Hamiltonian forms, for continuous and discrete mechanics, respectively. \S\ref{SystemNParticlesQInteraction} accounts for the introduction of the system under study, i.e.~$\mathcal{N}$ particles with quadratic interaction. We define the Continuous Damping Reduction (CDR), both in  Lagrangian and Hamiltonian forms, and prove in Theorem \ref{ContTheo} that the diagram Figure \ref{Sketch1} (left) is commutative. We illustrate the theorem by means of the Heat Bath example, where we show carefully the Damping Reduction process. In \S\ref{DiscCount} we establish the discrete counterpart of Theorem \ref{ContTheo}. Namely, we provide a continuous family of discretisation of the closed and open systems (depending on $\gamma\in[0,1]$), we define the discrete version of Damping Reduction (DDR) and prove in Theorem \ref{DiscTheo} that the diagram Figure \ref{Sketch1} (left) is commutative also at a discrete level. This accounts for the major contribution of the paper.  \S\ref{RubinSec} illustrates this last result with the example of transmission lines. We simulate the obtained integrators after reduction for $\gamma=0$ and a harmonic potential, observing a superior performance over the usual Euler discretisations.

\section{Continuous Lagrangian and Hamiltonian description of Closed/Open systems}\label{CLHDesc}

In the sequel we shall consider the configuration space of the studied systems as a finite dimensional smooth manifold $Q$. Moreover, $TQ$ and $T^*Q$ will denote its tangent and cotangent bundles, locally represented by coordinates $(q,\dot q)$ and $(q,p)$, respectively. For more details on the geometric formulation of mechanics we refer to \cite{AbMa}

\subsection{Closed systems}\label{ClosedCont}

Given a Lagrangian function $L:TQ\Flder\R$, the associated action functional in the time interval $[0,T]$ for a smooth curve $q:[0,T]\Flder Q$ is defined by $S(q)=\int_0^TL(q(t),\dot q(t))\,dt$.  Through Hamilton's principle, i.e. the true evolution of the system $q(t)$ with fixed endpoints $q(0)$ and $q(T)$ will satisfy

\begin{equation}\label{HPrinciple}
\delta\int_0^TL(q(t),\dot q(t))\,dt=0,
\end{equation}
we obtain the Euler-Lagrange equations via calculus of variations: 

\begin{equation}\label{ELeqs}
\frac{d}{dt}\lp\frac{\der L}{\der\dot q}\rp-\frac{\der L}{\der q}=0.
\end{equation}
Define the Legendre transformation: 
\begin{equation}\label{LegendreTrans}
\F L:TQ\Flder T^*Q; \quad (q,\dot q)\mapsto \left(q,p=\frac{\der L}{\der\dot q}\right).
\end{equation}
If  \eqref{LegendreTrans} is a global diffeomorphism we say that it is {\it hyperregular}, and we call the Lagrangian function {\it hyperregular}. Under the assumption of hyperregularity, through the Legendre transformation we can define the  Hamiltonian function $H:T^*Q\Flder\R$:
\begin{equation}\label{HamFunc}
H(q,p):=\bra p,\dot q\ket-L(q,\dot q),
\end{equation}
where $\bra\cdot,\cdot\ket:T^*Q\times TQ\Flder\R$ is the natural pairing. From the definition of the Hamiltonian function \eqref{HamFunc} it follows  that $L(q,\dot q)=\bra p,\dot q\ket-H(q,p)$. Furthermore, from \eqref{HPrinciple} we can write the stationary condition of the action functional in a Hamiltonian version, i.e.
\begin{equation}\label{HHPrin}
\delta\int_0^T\,\lc\bra p(t),\dot q(t)\ket-H(q(t),p(t))\rc\,dt=0.
\end{equation}
Again, using calculus of variations we obtain the Hamilton equations:

\begin{equation}\label{Heqs}
\dot q=\frac{\der H}{\der p},\quad\,\,\, \dot p=-\frac{\der H}{\der q}.
\end{equation}

We shall consider that the physical energy of the system is given by the Hamiltonian function \eqref{HamFunc}. It is easy to check that $dH(q,p)/dt=0$ under \eqref{Heqs}, showing that the system is closed. Equivalently, the Lagrangian energy
\begin{equation}\label{LagEner}
E(q,\dot q):=\Big<\frac{\der L}{\der\dot q},\dot q\Big>-L(q,\dot q),
\end{equation}
is invariant under \eqref{ELeqs}, i.e. $dE(q,\dot q)/dt=0.$
\subsection{Forced systems}

First we model the external forces (which might include damping, dragging, etc.) through the mapping
\begin{equation}\label{ExForces}
f_L:TQ\Flder T^*Q.
\end{equation}
The forced dynamics is provided by the Lagrange-d'Alembert principle \cite{Arnold, Bloch}: the true evolution of the system $q(t)$ between fixed points $q(0)$ and $q(T)$ will satisfy
\begin{equation}\label{LdAPrinL}
\delta\int_0^TL(q,\dot q)\,dt+\int_0^T\bra f_L(q,\dot q),\delta q\ket\,dt=0, 
\end{equation}
where $\delta q\in TQ$, which provides the forced Euler-Lagrange equations:
\begin{equation}\label{ForcedEL}
\frac{d}{dt}\lp\frac{\der L}{\der\dot q}\rp-\frac{\der L}{\der q}=f_L(q,\dot q).
\end{equation}
Now, the Lagrangian energy of the system \eqref{LagEner} is not preserved by \eqref{ForcedEL}.  In particular $dE(q,\dot q)/dt=\big< f_L(q,\dot q),\dot q\big>$, showing that this kind of systems is open.

The dual version of the Lagrange-d'Alembert principle \eqref{LdAPrinL} is naturally obtained through the Legendre transformation \eqref{LegendreTrans}. The dual external forces  $f_H:T^*Q\Flder T^*Q$ are defined by $f_H:=f_L\circ \lp\F L\rp^{-1}$ (we recall that we are assuming $L$ hyperregular), while the dynamics is established by
\begin{equation}\label{LdAPrinH}
\delta\int_0^T\lc\bra p,\dot q\ket-H(q,p)\rc\,dt+\int_0^T\bra f_H(q, p),\delta q\ket\,dt=0, 
\end{equation}
yielding the forced Hamilton equations 
\begin{equation}\label{ForcedHam}
\dot q=\frac{\der H}{\der p},\quad\,\,\, \dot p=-\frac{\der H}{\der q}+f_H(q,p).
\end{equation}
Obviously, the Hamiltonian function \eqref{HamFunc} is not preserved under \eqref{ForcedHam}. In particular $dH(q,p)/dt=\big<f_H(q,p),\frac{\der H}{\der p}\big>$.

\section{Discrete Lagrangian and Hamiltonian description of Closed/Open systems}\label{DiscreteFramework}

\subsection{Closed systems}

The construction of  the discrete version of mechanics relies on the substitution of $TQ$ by the Cartesian product $Q\times Q$ (note that these two spaces contain the same amount of information at local level) \cite{MaWe,MoVe}. The continuous curves $q(t)$ will be replaced by discrete ones, say $q_d=\lc q_k\rc_{0:N}:=\lc q_0,q_1,...,q_N\rc\in Q^{N+1}$, where $N\in \N$ and the power $N+1$ indicates the Cartesian product of $N+1$ copies of $Q$. Given an increasing sequence of times $\lc t_k=hk\,|\,k=0,...,N\rc\subset \R$, with $h=T/N$, the points in $q_d$ will be considered as an approximation of the continuous curve at time $t_k$, i.e. $q_k\simeq q(t_k).$ Defining the discrete Lagrngian $L_d:Q\times Q\Flder\R$ as an approximation of the action integral in one time step, say $L_d(q_k,q_{k+1})\simeq \int_{t_k}^{t_k+h}L(q(t),\dot q(t))\,dt$, we can establish the so called discrete action sum:
\begin{equation}\label{AcSum}
S(q_d)=\sum_{k=0}^{N-1}L_d(q_k,q_{k+1}).
\end{equation}
Applying the Hamilton's principle over \eqref{AcSum}, i.e. considering variations of $q_d$ with fixed endpoints $q_0, q_{N}$ and extremizing $S_d$, we obtain the discrete Euler-Lagrange equations
\begin{equation}\label{DEL}
D_1L_d(q_k,q_{k+1})+D_2L_d(q_{k-1},q_{k})=0,\quad k=1,...,N-1,
\end{equation}
where $D_1$ and $D_2$ denote the partial derivative with respect to the first and second variables, respectively. If $L_d$ is regular, i.e.~the matrix $\big[D_{12}L_d\big]$ is invertible, the equations \eqref{DEL} define a discrete Lagrangian flow $F_{L_d}:Q\times Q\Flder Q\times Q$; $(q_{k},q_{k+1})\mapsto (q_{k+1},q_{k+2})$, which  is normally called {\it variational integrator} of the continuous dynamics provided by the Euler-Lagrange equations \eqref{ELeqs} (indistinctly, we shall call the equations \eqref{DEL} also variational integrator).  Moreover,  \eqref{DEL} are a discretisation in finite differences of \eqref{ELeqs}.

In order to establish the Hamiltonian picture we need to introduce the discrete Legendre transforms.
From $L_d$ two of them can be defined:
\[
\F^{-}L_d,\F^{+}L_d:Q\times Q\Flder T^*Q,
\]
in particular
\begin{subequations}\label{DLT}
\begin{align}
\F^{-}L_d(q_k,q_{k+1})&=(\,\,\,q_k\,\,\,,p_k^{-}=-D_1L_d(q_k,q_{k+1})),\label{DLT:a}\\
\F^{+}L_d(q_k,q_{k+1})&=(q_{k+1},p_{k+1}^{+}=D_2L_d(q_k,q_{k+1})).\label{DLT:b}
\end{align}
\end{subequations}
We observe that the {\it momentum matching} condition, i.e.
\begin{equation}\label{MomMat}
p_k^{-}=p_{k}^{+},
\end{equation}
provides the discrete Euler-Lagrange equations \eqref{DEL} according to \eqref{DLT} (based on this, we shall refer indistinctly to the discrete Legendre transform as momentum matching). Under the regularity of $L_d$, both discrete Legendre transforms are invertible and the discrete Hamiltonian flow $\tilde F_{L_d}:T^*Q\Flder T^*Q$; $(q_k,p_{k})\mapsto (q_{k+1},p_{k+1})$ can be defined by any of the following identities:
\begin{equation}\label{DiscHamFlow}
\tilde F_{L_d}=\F^{+}L_d\circ (\F^{-}L_d)^{-1}=\F^{+}L_d\circ F_{L_d}\circ (\F^{+}L_d)^{-1}=\F^{-}L_d\circ F_{L_d}\circ (\F^{-}L_d)^{-1};
\end{equation}
see \cite{MaWe} for the proof. At the Hamiltonian level, the map $\tilde F_{L_d}$ is called {\it variational integrator} of the continuous dynamics provided by  the Hamilton equations \eqref{Heqs}. Moreover, the discrete equations provided by \eqref{DLT} are a discretisation in finite differences of \eqref{Heqs}.

A crucial feature of variational integrators is its {\it symplecticity}. If $\Omega_{T^*Q}$ is the canonical symplectic form on $T^*Q$ (which, according to Darboux theorem, can be locally written as $\Omega_{T^*Q}=dq\wedge dp$), define $\Omega_{Q\times Q}:=(\F^{-}L_d)^*\Omega_{T^*Q}=(\F^{+}L_d)^*\Omega_{T^*Q}$. Thus, the symplecticity of the variational integrators imply $F_{L_d}^*\Omega_{Q\times Q}=\Omega_{Q\times Q}$ \cite{MaWe, MoVe}, which furthermore imply that the energy cannot be conserved at the same time \cite{GeMa}. However, symplectic integrators have proven to present a stable energy behaviour even in long-term simulations \cite{SS}, behaviour that can be explained in terms of Backward Error Analysis \cite{BEA,HLW}.

\subsection{Forced systems}

As discrete version of the external forces \eqref{ExForces} we consider the maps
\[
f_{L_d}^{-}, f_{L_d}^{+}:Q\times Q\Flder T^*Q
\]
such that
\[
\bra f_{L_d}^{-}(q_k,q_{k+1}),\delta q_k\ket+\bra f_{L_d}^{+}(q_k,q_{k+1}),\delta q_{k+1}\ket\simeq \int_{t_k}^{t_k+h}\bra f_L(q,\dot q),\delta q\ket\,dt.
\]
Note that the previous equation implies that $f_{L_d}^{-}(q_k,q_{k+1})\in T^*_{q_k}Q$ and $f_{L_d}^{+}(q_k,q_{k+1})\in T^*_{q_{k+1}}Q$. The discrete Lagrange-d'Alembert principle \cite{MaWe,ObJuMa} provides discrete curves between fixed $q_0,q_{N}$ satisfying the critical condition
\[
\delta\sum_{k=0}^{N-1}L_d(q_k,q_{k+1})+\sum_{k=0}^{N-1}\Big[\bra f_{L_d}^{-}(q_k,q_{k+1}),\delta q_k\ket+\bra f_{L_d}^{+}(q_k,q_{k+1}),\delta q_{k+1}\ket\Big]=0.
\]
These curves are given by the forced discrete Euler-Lagrange equations
\begin{equation}\label{ForcedDEL}
D_1L_d(q_k,q_{k+1})+D_2L_d(q_{k-1},q_{k})+f_{L_d}^{-}(q_k,q_{k+1})+f_{L_d}^{+}(q_{k-1},q_{k})=0,\,\, k=1,...,N-1;
\end{equation}
they are a discretisation in finite differences of \eqref{ForcedEL} and, under the regularity of the matrix $\big[D_{12}L_d(q_k,q_{k+1})+D_2f_{L_d}^{-}(q_k,q_{k+1})\big]$, provide a forced discrete Lagrangian map $F_{L_d}^{\tiny \mbox{f}}:Q\times Q\Flder Q\times Q$ approximating their continuous solution. In the forced case, the discrete Legendre transformation is defined by
\begin{subequations}\label{DForcedLT}
\begin{align}
\F^{-}L_d^{\tiny \mbox{f}}(q_k,q_{k+1})&=(\,\,\,q_k\,\,\,,\,\,p_k^{-}=-D_1L_d(q_k,q_{k+1})-f_{L_d}^{-}(q_k,q_{k+1})),\label{DForcedLT:a}\\
\F^{+}L_d^{\tiny \mbox{f}}(q_k,q_{k+1})&=(q_{k+1},\,\,p_{k+1}^{+}=D_2L_d(q_k,q_{k+1})+f_{L_d}^{+}(q_k,q_{k+1})).\label{DForcedLT:b}
\end{align}
\end{subequations}
The momentum matching condition \eqref{MomMat} reproduces the forced discrete Euler-Lagrange equations \eqref{ForcedDEL}. Moreover, \eqref{DForcedLT} provide a discretisation in finite differences of \eqref{ForcedHam}, whereas \eqref{DiscHamFlow} (for $F_{L_d}^{\tiny  \mbox{f}}$ and $\F^{\pm}L_d^{\tiny  \mbox{f}}$) yields an approximation $\tilde F_{L_d}^{\tiny  \mbox{f}}:T^*Q\Flder T^*Q$ of their continuous flow.

\section{System of $\mathcal{N}$ particles with quadratic interaction}\label{SystemNParticlesQInteraction}

We shall consider a system of $\mathcal{N}$ particles. The configuration space of each particle is $Q=\R^d$. We will allow quadratic interaction among all of them, plus a conservative potential among the $n<\mathcal{N}$ particles of the open subsystem.

\subsection{Closed system}

Consider $(q_i,\dot q_i)\in T\R^d$  the coordinates and velocities of the $i$-th particle  for $i=1,...,\mathcal{N}$; $\,d,\mathcal{N}\in \N$. Define $(\mathfrak{q},\dot{\mathfrak{q}}):=((q_1,\dot q_1),(q_2,\dot q_2),...,(q_{\mathcal{N}},\dot q_{\mathcal{N}}))$ and the Lagrangian function  $L:(T\R^d)^{\mathcal{N}}\Flder\R$:
\begin{equation}\label{LagN}
L(\mathfrak{q},\dot{\mathfrak{q}})=\frac{1}{2}\sum_{i=1}^{\mathcal{N}}m_{i}\dot q_i^2-V(q_1,q_2,...,q_n)-\frac{1}{2}\sum_{i,j=1}^{\mathcal{N}}\lambda_{ij}q_iq_j,
\end{equation}
where $m_{i}\in\R^+$ for all $i$; $\lambda_{ij}\in\R$ for all $i,j$; $n\in\N$ with  $n\leq\mathcal{N}$ ($n$ the number of particles of the open subsystem) and $V:(\R^d)^n\Flder\R$ is a smooth function. As we shall see, different distributions of the particles in space correspond to different matrices $\Lambda:=(\lambda_{ij})\in\R^{\mathcal{N}\times\mathcal{N}}$. In all cases $\Lambda$ is symmetric, since it encodes the reciprocal interaction between $i,j$-th particles. Finally, for sake of simplicity, we denote $q_iq_j:=q_i^{T}q_j$  in terms of $\R^d$ scalar product; the same applies for velocities and momenta.
\begin{remark}
{\rm
Since $\Lambda$ is real and symmetric, it is diagonalisable by means of a particular rotation $R\in O(\mathcal{N})$. If we choose $m_i=m$ for all $i$ and $V=0$ in \eqref{LagN}, it is easy to see that a change of variables $r_i=R_{ij}q_j$ generates the new Lagrangian
\[
\hat L(\mathfrak{r},\dot{\mathfrak{r}})=\frac{1}{2}\sum_{i=1}^{\mathcal{N}}m\dot r_i^2-\frac{1}{2}\sum_{i=1}^{\mathcal{N}}\lambda_i r_i^2,
\]
with diag$(\lambda_1,...,\lambda_{\mathcal{N}})=R^T\Lambda R$. This Lagrangian acounts for a set of $\mathcal{N}$ decoupled harmonic oscillators, and therefore the interaction between the open subystem and the environment is absent. This exemplifies that the possibility of performing the damping reduction, as we shall detail in the examples, relies strongly on the particular expression of $\Lambda$, and that any transformation to a simpler problem might eliminate some of the physical properties that lead to the damping of the open subsystem.  
}\hfill $\diamond$
\end{remark}
The Euler-Lagrange \eqref{ELeqs} equations for the $i$-th particle are:
\begin{equation}\label{ELEqsN}
m_i\ddot q_i+\der_{q_i} V(q_1,...,q_n)+\sum_{j=1}^{\mathcal{N}}\lambda_{ij}q_j=0, \quad i=1,...,\mathcal{N}.
\end{equation}
It is apparent that $\der_{q_i}V(q_1,...,q_n)=0$ if $i> n$. The Legendre transformation \eqref{LegendreTrans} applied to \eqref{LagN} yields   $p_i=m_i\dot q_i$ for the $i$-th particle. Furthermore, it also provides the Hamiltonian function \eqref{HamFunc} $H:(T^*\R^d)^{\mathcal{N}}\Flder\R$ 
\begin{equation}\label{HamN}
H(\mathfrak{q},\mathfrak{p})=\sum_{i=1}^{\mathcal{N}}\frac{1}{2\,m_i} p_i^2+V(q_1,q_2..,q_n)+\frac{1}{2}\sum_{i,j=1}^{\mathcal{N}}\lambda_{ij}q_iq_j,
\end{equation}
where, for the $i$-th particle, the coordinates of $T^*\R^d$ are $(q_i,p_i)$ and we define $(\mathfrak{q},\mathfrak{p}):=((q_1,p_1),(q_2,p_2),$ $...,(q_{\mathcal{N}},p_{\mathcal{N}}))$. The Hamilton equations \eqref{Heqs} coming from \eqref{HamN} are
\begin{equation}\label{HamEqsN}
\dot q_i=\frac{1}{m_i}p_i,\quad\quad \dot p_i=-\der_{q_i} V(q_1,...,q_n)-\sum_{j=1}^{\mathcal{N}}\lambda_{ij}q_j, 
\end{equation}
for $i=1,...,\mathcal{N}.$ 

Since we are dealing with closed systems, the energy, both in its Lagrangian \eqref{LagEner} and Hamiltonian forms
\[
\begin{split}
E(\mathfrak{q},\dot{\mathfrak{q}})&=\sum_{i=1}^{\mathcal{N}}\frac{1}{2}m_{i}\dot q_i^2+V(q_1,...,q_n)+\frac{1}{2}\sum_{i,j=1}^{\mathcal{N}}\lambda_{ij}q_iq_j,\\
H(\mathfrak{q},\mathfrak{p})&=\sum_{i=1}^{\mathcal{N}}\frac{1}{2\,m_i} p_i^2+V(q_1,...,q_n)+\frac{1}{2}\sum_{i,j=1}^{\mathcal{N}}\lambda_{ij}q_iq_j,
\end{split}
\]
is invariant under the dynamics \eqref{ELEqsN} and \eqref{HamEqsN}, respectively. This is $dE(\mathfrak{q},\dot{\mathfrak{q}})/dt=dH(\mathfrak{q},\mathfrak{p})/dt=0,$ which can be checked after a straightforward computation, as discussed in \S\ref{ClosedCont}.

\subsection{Continuous Damping Reduction: the open system}

By Continuous Damping Reduction (CDR henceforth) of the Euler-Lagrange\eqref{ELEqsN}/Hamilton\eqref{HamEqsN} dynamical equations we mean the process of focusing on a subsystem of $n\in\N$ particles ($n<N$) which is linearly damped in all coordinates due to the physical interaction with the enviroment. The CDR, which has its physical roots in the quadratic interaction
\[
\frac{1}{2}\sum_{i,j=1}^{\mathcal{N}}\lambda_{ij}q_iq_j,
\]
will obey a particular physical process depending on the kind of system we are dealing with. Let us define $(\tilde{\mathfrak{q}},\dot{\tilde{\mathfrak{q}}}):=((q_1,\dot q_1),...,(q_n,\dot q_n))$ and $(\tilde{\mathfrak{q}},\tilde{\mathfrak{p}}):=((q_1,p_1),...,(q_n,p_n))$, the coordinates, velocities and momenta of the open subsystem. Moreover, consider its Lagrangian $L^s:(T\R^d)^n\Flder\R$ and Hamiltonian  $H^s:(T^*\R^d)^n\Flder\R$ functions 
\begin{subequations}\label{Subsystem}
\begin{align}
L^s(\tilde{\mathfrak{q}},\dot{\tilde{\mathfrak{q}}})&=\sum_{a=1}^n\frac{1}{2}m_a\dot q_a^2-V(q_1,...,q_n), \label{Subsystem:L}\\
H^s(\tilde{\mathfrak{q}},\tilde{\mathfrak{p}})&=\sum_{a=1}^n\frac{1}{2m_a} p_a^2+V(q_1,...,q_n),\label{Subsystem:H}
\end{align}
\end{subequations}
where the superscript $s$ stands for {\it subsystem}. It is easy to check that the Legendre transformation $\F L^s$ \eqref{LegendreTrans}  relating \eqref{Subsystem:L} and \eqref{Subsystem:H} is locally defined  by $p_a=\F L^s(q_a,\dot q_a)=m_a\dot q^a$. Given this, we define the CDR as follows

\begin{definition}\label{CDRDef}
For $i,j=1,...,\mathcal{N}$, and $a,b=1,...,n$; we define the {\rm CDR} in its Lagrangian and Hamiltonian forms, respectively, as
\begin{subequations}
\begin{align}
\sum_{j=1}^{\mathcal{N}}\lambda_{ij}q_j&\longrightarrow -\sum_{b=1}^n\tilde\lambda_{ab}\dot q_b, \label{LagDR}\\
\sum_{j=1}^{\mathcal{N}}\lambda_{ij}q_j&\longrightarrow -\sum_{b=1}^n\tilde\lambda_{ab}\frac{p_b}{m_b}\label{HamDR},
\end{align}
\end{subequations}
where  \eqref{HamDR} follows directly from \eqref{LagDR} just employing the Legendre transform.
\end{definition}
Note that, in order to emphasize that after CDR we are going to focus in the open subsystem, we employ different indices, particularly $i,j$ for the closed system and $a,b$ for the open one. Finally, the relationship between the matrices $\Lambda$ and $\tilde\Lambda:=(\tilde\lambda_{ab})\in \R^{n\times n}$ (corresponding to the distribution of the open subsystem) is determined by the particular physical process producing the CDR.

The Lagrangian CDR \eqref{LagDR} applied to \eqref{ELEqsN} provides the dynamical equations of the damped open subsystem, i.e.
\begin{equation}\label{LagAfterDR}
m_a\ddot q_a+\der_{q_a} V(\tilde{\mathfrak{q}})=\sum_{b=1}^n\tilde\lambda_{ab}\dot q_b.
\end{equation}
Define the external forces by
\begin{subequations}\label{Forces}
\begin{align}
f_L(\tilde{\mathfrak{q}},\dot{\tilde{\mathfrak{q}}})_a&=\,\,\,\,\sum_{b=1}^n\tilde\lambda_{ab}\dot q_b,
 \label{Forces:L}\\
f_H(\tilde{\mathfrak{q}},\tilde{\mathfrak{p}})_a&=\sum_{b=1}^n\tilde\lambda_{ab}p_b/m_b,\label{Forces:H}
\end{align}
\end{subequations}
again for $a=1,...,n$; linked through $\F L^s$. It is easy to check that the Lagrange-d'Alembert principle \eqref{LdAPrinL} for $L^s(\tilde{\mathfrak{q}},\dot{\tilde{\mathfrak{q}}})$ \eqref{Subsystem:L} and external forces $f_L(\tilde{\mathfrak{q}},\dot{\tilde{\mathfrak{q}}})$ \eqref{Forces:L} provides the open Lagrangian dynamics \eqref{LagAfterDR}.

The previous discussion allows us to establish the following theorem.

\begin{theorem}\label{ContTheo}
Given a Lagrangian function $L(\mathfrak{q},\dot{\mathfrak{q}})$ \eqref{LagN} for a closed system of $\mathcal{N}$ interacting particles, the Lagrangian $L^s(\tilde{\mathfrak{q}},\dot{\tilde{\mathfrak{q}}})$ of the subsytem of $n$ particles \eqref{Subsystem:L}, the external forces \eqref{Forces} and the {\rm CDR} given in Definition \ref{CDRDef}, the following diagram is commutative.

{\rm \begin{equation}\label{DiagramCont}
\xymatrix{
*[F]{\begin{array}{c}
\\
L(\mathfrak{q},\dot{\mathfrak{q}})\,\,\,\eqref{LagN}
\\\\
\end{array}}\ar[r]^{\F L}\ar[d]_{\mbox{HP} \,\eqref{HPrinciple}}  &  *[F]{\begin{array}{c}
\\
H(\mathfrak{q},\mathfrak{p})\,\,\,\eqref{HamN}
\\\\
\end{array}}\ar[d]^{\mbox{HP}\,\eqref{HHPrin}} \\
*[F]{\begin{array}{c}
\\
m_i\ddot q_i+\der_{q_i}V(\tilde{\mathfrak{q}})+\sum_{j=1}^{\mathcal{N}}\lambda_{ij}q_j=0
\\\\
\end{array}}\ar[d]_{\mbox{CDR}\,\eqref{LagDR}} & *[F]{\begin{array}{c}
\\
\dot q_i=\frac{1}{m_i}p_i,\,\, \dot p_i=-\der_{q_i} V(\tilde{\mathfrak{q}})-\sum_{j=1}^{\mathcal{N}}\lambda_{ij}q_j
\\\\
\end{array}}\ar[ddd]^{\mbox{CDR}\,\eqref{HamDR}}\\
*[F]{\begin{array}{c}
\\
m_a\ddot q_a+\der_{q_a} V(\tilde{\mathfrak{q}})=\sum_{b=1}^n\tilde\lambda_{ab}\dot q_b
\\\\
\end{array}}\ar@{<=>}[d]_{\mbox{LdA}} &  \\
*[F]{\begin{array}{c}
\\
\delta\int_0^TL^s(\tilde{\mathfrak{q}},\dot{\tilde{\mathfrak{q}}})\,dt+\int_0^T\bra f_L(\tilde{\mathfrak{q}},\dot{\tilde{\mathfrak{q}}}),\delta \tilde{\mathfrak{q}}\ket\,dt=0\\
\eqref{Subsystem:L}\quad\quad\quad\quad\quad\eqref{Forces:L} \\
\end{array}}\ar[d]_{\F L^s} & \\
*[F]{\begin{array}{c}
\\
\delta\int_0^T\lc\bra\tilde{\mathfrak{p}},\dot{\tilde{\mathfrak{q}}}\ket-H^s(\tilde{\mathfrak{q}},\tilde{\mathfrak{p}})\rc\,dt+\int_0^T\bra f_H(\tilde{\mathfrak{q}},\tilde{\mathfrak{p}}),\delta \tilde{\mathfrak{q}}\ket\,dt=0 
\\\eqref{Subsystem:H}\quad\quad\quad\quad\quad\eqref{Forces:H}\\
\end{array}}\ar@/_4pc/[r]_{\mbox{LdA}} &  *[F]{\begin{array}{c}
\\
\mbox{FHEqs}_n
\\\\
\end{array}}
}
\end{equation}}
In other words, its two branches, i.e.
{\rm\begin{itemize}
\item[1.] LdA(Ham) $\circ\,\F L^s\,\circ$ LdA(Lag) $\circ$ CDR(Lag) $\circ$ HP(Lag)

\item[2.] CDR(Ham) $\circ$  HP(Ham) $\circ$ $\F L$,
\end{itemize}}
\hspace{-0.55cm} provide  the same set of Forced Hamilton Equations  for the open subsystem of $n$ linearly damped particles (which we denote {\rm FHEqs$_n$}).
\end{theorem}

\begin{proof}
Applying the Hamiltonian CDR \eqref{HamDR} over \eqref{HamEqsN} leads to
\begin{equation}\label{HamAfterDR}
\dot q_a=\frac{p_a}{m_a},\quad\quad \dot p_a=-\der_{q_a}V(\tilde{\mathfrak{q}})+\sum_{b=1}^n\tilde\lambda_{ab}\frac{p_b}{m_b},
\end{equation}
which are FHEqs$_n$. This accounts for branch 2 in diagram \eqref{DiagramCont}.

On the other hand, as metioned above the forced Euler-Lagrange equations for the damped subsystem of $n$ particles \eqref{LagAfterDR} can be obtained through the Lagrange-d'Alembert principle for $L^s(\tilde{\mathfrak{q}},\dot{\tilde{\mathfrak{q}}})$ \eqref{Subsystem:L} and external forces $f_L(\tilde{\mathfrak{q}},\dot{\tilde{\mathfrak{q}}})$ \eqref{Forces:L}. The application of the Legendre transform over this principle yields its Hamiltonian expression for $H^s(\tilde{\mathfrak{q}},\tilde{\mathfrak{p}})$ \eqref{Subsystem:H} and external forces $f_H(\tilde{\mathfrak{q}},\tilde{\mathfrak{p}})$ \eqref{Forces:H}, which in coordinates looks like
\[
\delta\int_0^T\Big\{\sum_{b=1}^n\Big( p_b\dot q_b-\frac{p_b^2}{2m_b}\Big)-V(\tilde{\mathfrak{q}})\Big\}\,dt-\int_0^T\Big(\sum_{b=1}^n\tilde\lambda_{ab}\frac{p_b}{m_b}\Big)\,\delta q_a\,dt=0.
\]
Taking variations in the first integral, doing integration by parts and considering that $\delta q_a(0)=\delta q_a(T)=\delta p_a(0)=\delta p_a(T)=0$ yields directly FHEqs$_n$. This accounts for the branch 1 and proves the theorem.
\end{proof}

\begin{remark}
{\rm Theorem \ref{ContTheo} does not necessarily apply to systems with the kinetic energy of the form $T=\frac{1}{2}\sum_{ij}\kappa_{ij}\dot q_i\dot q_j$, where the matrix $\mathcal{K}=(\kappa_{ij})$ is not diagonal. Such terms correspond to mechanical systems known as inerters \cite{DavidsPaper,Smith}, for which the kinetic energy is defined as:
\[
T = \frac{1}{2}b(\dot{q}_{\tiny\mbox{fin}} - \dot{q}_{\tiny\mbox{in}})^2
\]
where $\dot{q}_{\tiny\mbox{fin}}$ and $\dot{q}_{\tiny\mbox{in}}$ denote the terminal points of an inerter, and $b$ denotes the inertance. In order for the diagram to commute for such systems one needs to apply linear transformations on the coordinates and momenta.  In order to illustrate this phenomenon, let us  consider two masses ($m_{\tiny\mbox{in}}$, $m_{\tiny\mbox{fin}}$) coupled to one another through a transmission line comprising a set of inerters and springs  (with $\mathfrak{n}\in\N$ inerters).  After applying the Legendre transformation to the Lagrangian (see \cite{DavidsPaper}), the Hamiltonian function of the closed system reads
\[
\begin{split}
H(\mathfrak{q},\mathfrak{p})=&\,\,\frac{1}{2m_{\tiny\mbox{in}}}p_{\tiny\mbox{in}}^2+\frac{1}{2m_{\tiny\mbox{fin}}}p_{\tiny\mbox{fin}}^2+\frac{1}{2b}\sum_{i=1}^{\mathfrak{n}}p_i^2
+ \frac{1}{m_{\tiny\mbox{fin}}}p_{\tiny\mbox{fin}}\sum_{i}^{\mathfrak{n}}p_i+\frac{1}{2m_{\tiny\mbox{fin}}}\sum_{i,j=1}^{\mathfrak{n}}p_ip_j\\
&\quad\quad\quad+\frac{1}{2}k_{\tiny\mbox{in}}q_{\tiny\mbox{in}}^2+\frac{1}{2}k_{\tiny\mbox{fin}}q_{\tiny\mbox{fin}}^2+\frac{1}{2}k(q_{\mathfrak{n}}-q_{\tiny\mbox{fin}})^2
+\frac{k}{2}\sum_{i=1}^{\mathfrak{n}}(q_i-q_{i-1})^2,
\end{split}
\]
where $k_{\tiny\mbox{in}},\,k_{\tiny\mbox{fin}},\,k\in\R$ are spring coupling constants  between the first mass and the first interter, between the second mass and the final inerter, and between the inerters themselves, respectively.  We observe a non-diagonal kinetic energy due to the coupling among the momenta  within the inerter line, i.e.  $\,\,p_i\,p_j$, and the inerters and final mass, i.e.~$p_{\tiny\mbox{fin}}\,p_i$. It can be shown that, after Hamiltonian CDR (similar to \S\ref{RubinSec}, see \cite{DavidsPaper} for more details), the equations for the open system read 
\[
\begin{split}
m_{\tiny\mbox{in}}:&\quad \dot q_{\tiny\mbox{in}}=\frac{p_{\tiny\mbox{in}}}{m_{\tiny\mbox{in}}},\hspace{2.7cm}\,\,\dot p_{\tiny\mbox{in}}=-k_{\tiny\mbox{in}}q_{\tiny\mbox{in}}-D\frac{p_{\tiny\mbox{in}}}{m_{\tiny\mbox{in}}}-D\frac{1}{m_{\tiny\mbox{fin}}}\Big( p_{\tiny\mbox{fin}}+\sum_{i=1}^{\mathfrak{n}}p_i\Big),\\
m_{\tiny\mbox{fin}}:&\quad \dot q_{\tiny\mbox{fin}}=\frac{1}{m_{\tiny\mbox{fin}}}\Big(p_{\tiny\mbox{fin}}+\sum_{i=1}^{\mathfrak{n}}p_i\Big),\quad\,\, \dot p_{\tiny\mbox{fin}}=-k_{\tiny\mbox{fin}}q_{\tiny\mbox{fin}}-\frac{D}{m_{\tiny\mbox{fin}}}\Big(p_{\tiny\mbox{fin}}+\sum_{i=1}^{\mathfrak{n}}p_i\Big)+D\frac{p_{\tiny\mbox{in}}}{m_{\tiny\mbox{in}}}\\
&\hspace{8.5cm}-\sum_{i=1}^{\mathfrak{n}}k(q_{i+1}-2q_i+q_{i-1}),
\end{split}
\]
where $D$ depends on the $k'$s and b. It is apparent that these are not linearly damped equations. However, it can be shown as well that redefining the momenta by $\tilde p_{\tiny\mbox{fin}}:=m_{\tiny\mbox{fin}}\dot q_{\tiny\mbox{fin}}=p_{\tiny\mbox{fin}}+\sum_{i=1}^{\mathfrak{n}}p_i$ the open equations become
\[
\begin{split}
m_{\tiny\mbox{in}}:&\quad\quad \dot q_{\tiny\mbox{in}}=\frac{p_{\tiny\mbox{in}}}{m_{\tiny\mbox{in}}},\hspace{2cm}\,\,\,\,\dot p_{\tiny\mbox{in}}=-k_{\tiny\mbox{in}}q_{\tiny\mbox{in}}-D\lp\frac{p_{\tiny\mbox{in}}}{m_{\tiny\mbox{in}}}-\frac{\tilde p_{\tiny\mbox{fin}}}{m{\tiny\mbox{fin}}}\rp,\\
m_{\tiny\mbox{fin}}:&\quad\quad \dot q_{\tiny\mbox{fin}}=\frac{\tilde p_{\tiny\mbox{fin}}}{m_{\tiny\mbox{fin}}},\hspace{2cm}\,\,\dot{\tilde p}_{\tiny\mbox{fin}}=-k_{\tiny\mbox{fin}}q_{\tiny\mbox{fin}}-D\lp\frac{\tilde p_{\tiny\mbox{fin}}}{m_{\tiny\mbox{fin}}}-\frac{p_{\tiny\mbox{in}}}{m{\tiny\mbox{in}}}\rp,
\end{split}
\]
which are \eqref{HamAfterDR} for two particles with $V(q_{\tiny\mbox{in}},q_{\tiny\mbox{fin}})=k_{\tiny\mbox{in}}q_{\tiny\mbox{in}}^2/2+k_{\tiny\mbox{fin}}q_{\tiny\mbox{fin}}^2/2$ and $\tilde\Lambda=\begin{bmatrix} -D&D\\D&-D\end{bmatrix}$. 
} \hfill $\diamond$ 
\end{remark}

\subsection{Example: Heat Bath}\label{HBSection}

\begin{figure}[!htb]
\includegraphics[scale=0.25]{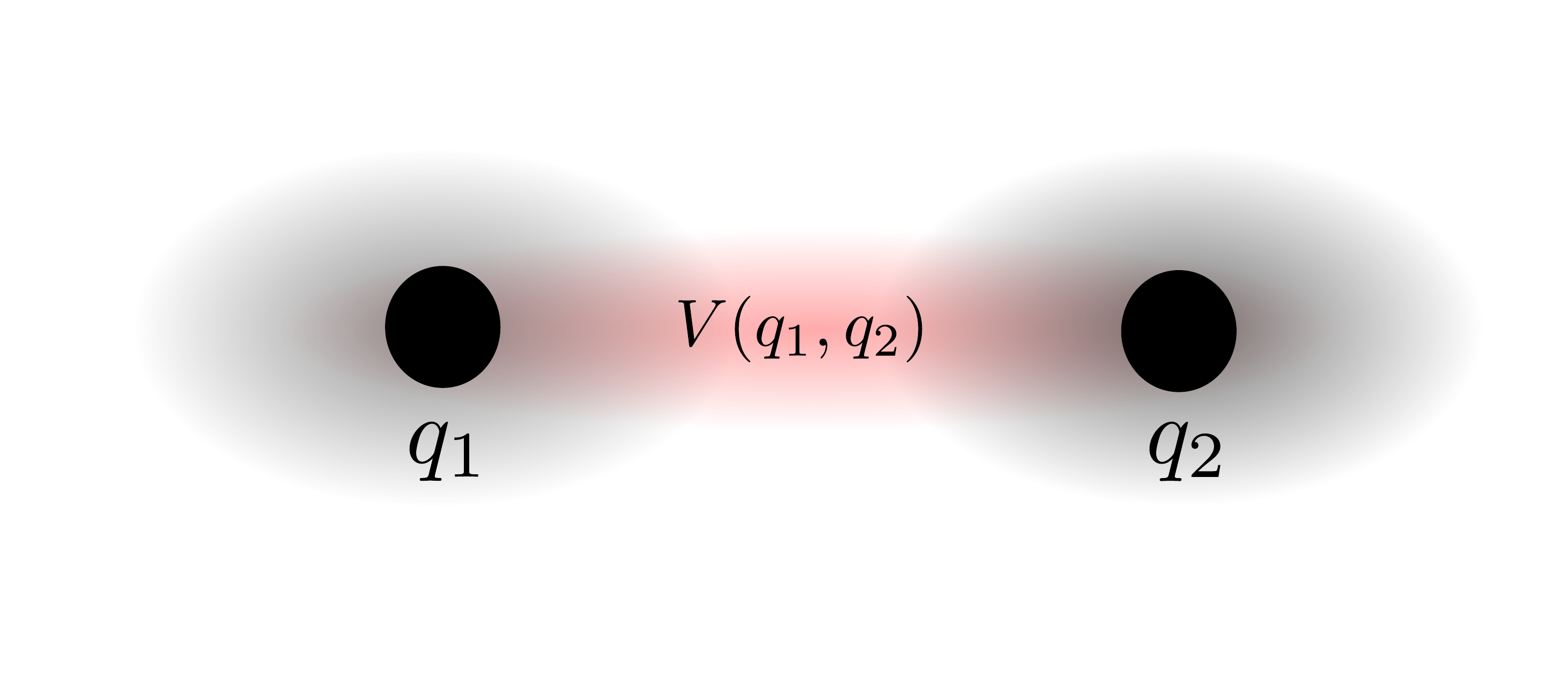}
\caption{Two systems immersed within their linear dissipative environment coupled through an arbitrary interaction potential.}
\label{fig1}
\end{figure}

In the example sections (both here and $\S$\ref{RubinSec}) we consider a set of two linear dissipative systems that are interacting with one another through an arbitrary interaction potential as illustrated in Figure \ref{fig1}.

\begin{figure}[!htb]
\includegraphics[scale=0.3]{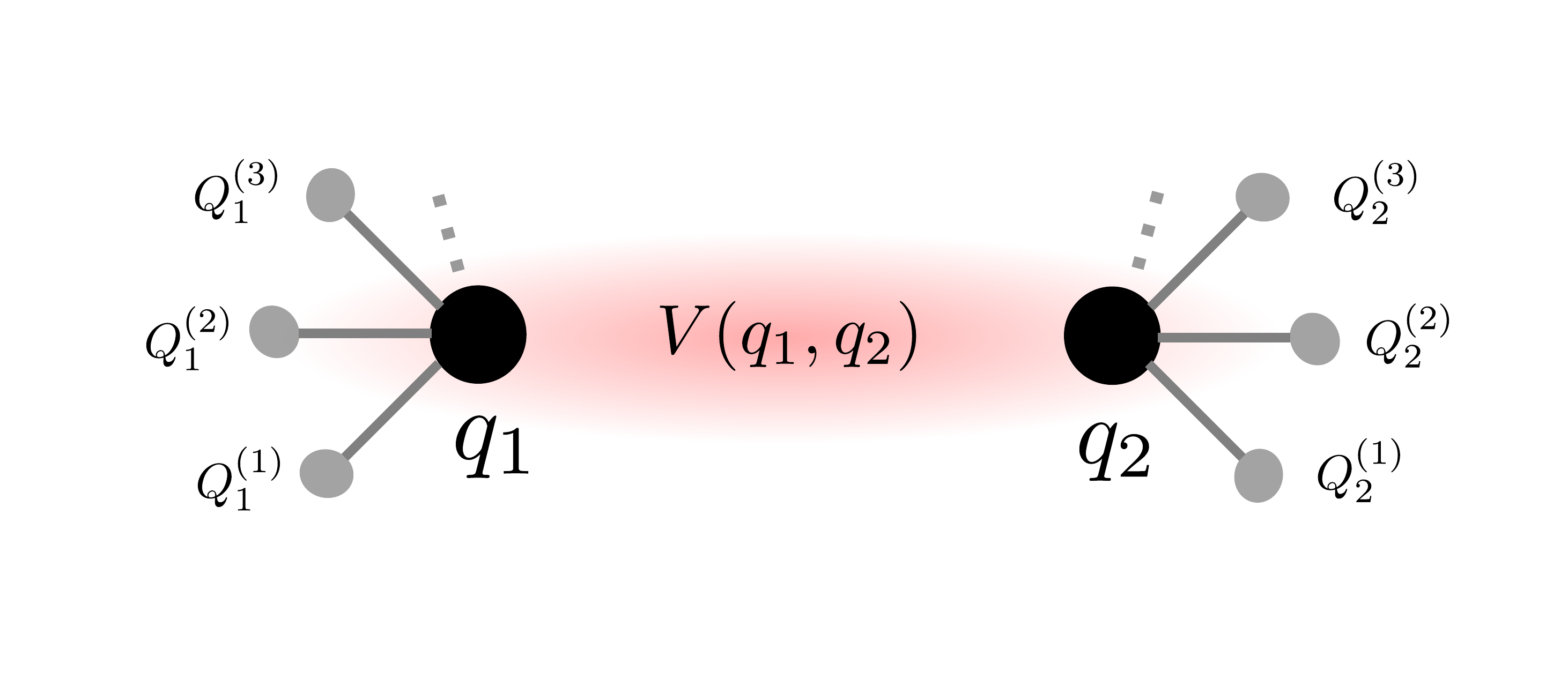}
\caption{Heat bath simulated in terms of a set of harmonic oscillators coupled independently to the open system.}
\label{heatbath}
\end{figure}

A possible approach in modeling the linear dissipative environment is illustrated in Figure \ref{heatbath}, i.e. to simulate the environment with a set of harmonic oscillators that are independently coupled in a bilinear form to the open system {\footnote{This model is referred to as the Caldeira-Leggett model \cite{CL, Zwanzig}. }}. This model has been widely used in studying statistical mechanical systems both in classical and quantum dynamics.

We observe that the open subsystem contains two particles, this is $n=2$, corresponding to $q_1$ and $q_2$. The total set of particles is denoted by $\lc q_i\rc_{1}^{\mathcal{N}}:=\lc q_1,q_2,...,q_{\mathcal{N}-1},q_{\mathcal{N}}\rc$. According to \eqref{LagDR} we are going to employ the index $a=1,2,$ to denote the particles in the open subsystem. Moreover, for simplicity, we consider that both heat baths contain the same amount of particles, i.e. $(\mathcal{N}-2)/2$. Taking this into account, we can rearrange the particle ordering as
\begin{equation}\label{Notation}
\begin{split}
\lc q_i\rc_{1}^{\mathcal{N}}&:=\lc q_1,q_2,\lc Q_1^j\rc_{1}^{(\mathcal{N}-2)/2},\lc Q_2^j\rc_{1}^{(\mathcal{N}-2)/2}\rc,\\
\lc Q_1^j\rc_{1}^{(\mathcal{N}-2)/2}&:=\lc q_3,...,q_{(\mathcal{N}-2)/2}\rc,\\
\lc Q_2^j\rc_{1}^{(\mathcal{N}-2)/2}&:=\lc q_{(\mathcal{N}-2)/2+1},...,q_{\mathcal{N}}\rc,
\end{split}
\end{equation}
where naturally $Q_a^j$ denotes the $j$-th oscillator coupled to the $a$-th open system (not to confuse with $Q$ the smooth manifold introduced in \S\ref{CLHDesc}). Using this notation, the Lagrangian dynamics of the closed system is given by
\begin{equation}\label{LagFuncClosed}
\begin{split}
&L(\mathfrak{q},\dot{\mathfrak{q}})=\frac{1}{2}\sum_{a=1}^2m_a\dot q_a^2+\frac{1}{2}\sum_{a=1}^2\sum_{j=1}^{(\mathcal{N}-2)/2}m_a^j(\dot Q_a^j)^2\\
&\quad\quad\quad\,\,\,\,\,\,-V(q_1,q_2)-\frac{1}{2}\sum_{a=1}^2\sum_{j=1}^{(\mathcal{N}-2)/2}\omega_a^j(Q_a^j - \frac{\gamma_a^j}{(\omega_a^j)^2}q_a)^2.
\end{split}
\end{equation}
Here, $\omega_a^j$ denotes the frequency of the $j$-th oscillator coupled to the $a$-th open system, and $\gamma_a^j$ denotes the strength of the interaction of the $j$-th oscillator and the $a$-th open system; henceforth,  we shall consider that $m_a^j=M$ for all $a$ and $j$. We can rearrange \eqref{LagFuncClosed} to obtain $L(\mathfrak{q},\dot{\mathfrak{q}}) = \sum_{a =1}^2 L^s (q_a,\dot q_a) + \sum_{a=1}^2 L^{\tiny\mbox{hb}} (q_a,\dot q_a ; Q_a,\dot Q_a)$,  where $L^s$ and $L^{\tiny\mbox{hb}}$ encode the open system dynamics and the heat bath including its interaction with the open system, respectively:
\begin{subequations}\label{LagSHB}
\begin{align}
L^s (q_a,\dot q_a) &= \,\,\,\,\,\,\frac{1}{2}m_a\dot q_a^2 - V(q_1,q_2), \label{LagSHB:a}\\
L^{\tiny\mbox{hb}}(q_a,\dot q_a ; Q_a,\dot Q_a) &= \sum_{j=1}^{(\mathcal{N}-2)/2}\bigg( \frac{1}{2}M(\dot Q_a^j)^2-\frac{1}{2}\omega_a^j(Q_a^j - \frac{\gamma_a^j}{(\omega_a^j)^2}q_a)^2\bigg).
\end{align}
\end{subequations}

Using \eqref{Notation}, \eqref{LagFuncClosed} and comparing to Figure \ref{heatbath}, we observe that the quadratic interaction matrix $\Lambda$ is given by
\begin{small}
\[
2\Lambda=\begin{bmatrix}
\sum_j(\gamma_1^j)^2/(\omega_1^j)^3 & 0& \gamma_1^1/\omega_1^1 & \cdots& \gamma_1^{(\mathcal{N}-2)/2}/\omega_1^{(\mathcal{N}-2)/2} &0 &\cdots &0\\
0 & \sum_j(\gamma_2^j)^2/(\omega_2^j)^3& 0 &\cdots &0 & \gamma_2^1/\omega_2^1 &\cdots & \gamma_2^{(\mathcal{N}-2)/2}/\omega_2^{(\mathcal{N}-2)/2}\\
\gamma_1^1/\omega_1^1&0& \omega_1^1 & \cdots& 0 &0 & \cdots& 0\\
\vdots & \vdots&\vdots & \ddots&\vdots & \vdots& & \vdots\\
\gamma_1^{(\mathcal{N}-2)/2}/\omega_1^{(\mathcal{N}-2)/2} & 0& 0&\cdots &\omega_1^{(\mathcal{N}-2)/2} & 0& \cdots&0\\
 0&\gamma_2^1/\omega_2^1 & 0&\cdots &0 &\omega_2^1 &\cdots &0\\
 \vdots &\vdots & \vdots& & \vdots& \vdots& \ddots&\vdots\\
 0& \gamma_2^{(\mathcal{N}-2)/2}/\omega_2^{(\mathcal{N}-2)/2}& 0&\cdots &0 &0 &\cdots &\omega_2^{(\mathcal{N}-2)/2}
\end{bmatrix}.
\]
\end{small}
Using the Legendre transformation \eqref{LegendreTrans} on \eqref{LagFuncClosed}, we obtain the Hamiltonian function $ H(\mathfrak{q},\mathfrak{p}) = \sum_{a =1}^2 H^s (q_a,p_a) + \sum_{a=1}^2 H^{\tiny\mbox{hb}} (q_a,p_a ; Q_a,P_a),$ where:
\begin{subequations}\label{HamSHB}
\begin{align}
H^s (q_a,p_a) &= \frac{p_a^2}{2 m_a} + V(q_1,q_2), \label{HamSHB:a}\\
H^{\tiny\mbox{hb}}(q_a,p_a ; Q_a,P_a) &= \sum_{j=1}^{(\mathcal{N}-2)/2}\bigg( \frac{(P_a^j)^2}{2M}+\frac{1}{2}\omega_a^j(Q_a^j - \frac{\gamma_a^j}{(\omega_a^j)^2}q_a)^2\bigg).
\end{align}
\end{subequations}
The dynamics of the closed system can be obtained by applying the Hamilton's equation \eqref{Heqs} to \eqref{HamSHB}, yielding:
\begin{subequations}
\begin{align}
\dot{q}_a &= \frac{p_a}{m_a},\\
\dot{p}_a &= -\der_{q_a}V + \sum_{j=1}^{(\mathcal{N}-2)/2} \gamma^j_a(Q_a^j - \frac{\gamma_a^j}{(\omega_a^j)^2}q_a), \\
\dot{Q}_a^j &= \frac{P_a^j}{M},\label{hamiltonEqc}\\
\dot{P}_a^j &= -(\omega_a^j)^2Q_a^j + \gamma_a^j q_a.
\label{hamiltonEqd}
\end{align}
\end{subequations}
The heat bath degrees of freedom can be eliminated by solving their dynamics \eqref{hamiltonEqc}, \eqref{hamiltonEqd} in terms of the open system degrees of freedom: 
\[
Q_a^j(t) = Q_a^j(0)\cos (\omega_a^j t) + P_a^j(0)\sin (\omega_a^j t)/\omega_a^j + \gamma_a^j \int_0^t dt^{\prime} q_a(t^{\prime})\frac{\sin (\omega_a^j(t - t^{\prime}))}{\omega_a^j}.
\]
Doing integration by parts and using the solution for the open system in equation (\ref{hamiltonEqd}), one obtains the following form of the solution for the open system: 
\begin{subequations}\label{HamOpen}
\begin{align}
\dot{q}_a &= \frac{p_a}{m_a} \label{Langevin0},\\
\dot{p}_a &= -\der_{q_a}V - \int _0^t dt^{\prime} \gamma_a(t - t^{\prime})\dot{q}_a + F(t).
\label{Langevin1}
\end{align}
\end{subequations} 
Here $\gamma_a(t - t^{\prime})$ denotes the memory function \cite{Zwanzig} for the dissipation. In other words, the dissipative dynamics is not local-in-time: 
\begin{equation}
\gamma_a(t - t^{\prime}) = \sum_{j=1}^{(\mathcal{N}-2)/2} \frac{(\gamma_a^j)^2}{(\omega_a^j)^2}\cos (\omega_a^j (t - t^{\prime})).
\label{Kernel}
\end{equation}
As the system dynamics becomes local-in-time the memory function converges to the delta function, i.e. $\gamma_a(t-t^{\prime}) \Flder \gamma_a\delta(t - t^{\prime})$ with $\gamma_a\in\R$. This can be understood by observing that replacing the $\gamma_a(t-t^{\prime})$ with  $\gamma_a\delta(t - t^{\prime})$ turns the integro differential equation (\ref{Langevin1}) in to local differential equation.  An environment with a memory function proportional to a delta function is referred to as a Markovian or memoryless environment. Typically, an environment with large number of degrees of freedom exhibit Markovian behavior. In the equation \eqref{Kernel}, as $\mathcal{N}\rightarrow \infty$ the sum can be replaced with a integral and by choosing the right distribution for the coupling constants, e.g.~$\gamma^j = \gamma \omega^j$:
\begin{equation}
\gamma_a(t - t^{\prime}) = \gamma_a \int_0^{\infty} d\omega_a \cos (\omega_a (t - t^{\prime}))\propto \delta (t - t^{\prime}).
\end{equation}  
$F(t)$ is the force term which depends on the initial conditions of the heat bath and the initial position of the open system:
\[
F(t) = \sum_{j=1}^{(\mathcal{N}-2)/2} \gamma_a^j P_a^j(0)\sin (\omega_a^j t)/\omega_a^j + \sum_{j=1}^{(\mathcal{N}-2)/2}  \gamma^j_a \Big( Q_a^j(0)-\frac{\gamma_a^j}{(\omega^j_a)^2}q_a(0)\Big)\cos (\omega_a^j t).
\]
Here $F(t)$ can be considered as the random fluctuation force exerted on the open system by the environment.

In such case, the equations of the open system \eqref{HamOpen} become
\begin{subequations}\label{HamOpenFinal}
\begin{align}
\dot{q}_a &= \frac{p_a}{m_a},\\
\dot{p}_a &= -\der_{q_a}V - \gamma_a\dot{q}_a + F(t)=-\der_{q_a}V - \frac{\gamma_a}{m_a}p_a + F(t),
\end{align}
\end{subequations} 
which correspond to equations \eqref{HamAfterDR}. In conclusion, the physical process leading to equations \eqref{HamOpenFinal} represents the Hamiltonian Continuous Damping Reduction (CDR) \eqref{HamDR} and the branch 2 in Theorem \ref{ContTheo} (also the process scketched in Figure \ref{Sketch1} (right)). Observe that equations \eqref{HamOpenFinal} contain the non-autonomous term $F(t)$, that only vanishes if $P_a^j(0)=0$ and $Q_a^j(0)=\gamma_a^jq_a(0)/(\omega_a^j)^2$, which is a rather unnatural scenario. Then, apparently this example does not fit diagram \eqref{DiagramCont}. However, it is easy to see that the diagram admits a non-autonomous extension, which consists basically in adding external non-autonomous forces after damping reduction and also in $f_L,\,f_H.$

Furthermore, the damping matrix $\tilde\Lambda$ is given by
\begin{equation}\label{DampedMatrix}
\tilde\Lambda=\begin{bmatrix}
\gamma_1&0\\
0&\gamma_2
\end{bmatrix}.
\end{equation}

Branch 1 in Theorem \ref{ContTheo} is given by the following steps. First we obtain the Euler-Lagrange equations from Lagrangian \eqref{LagFuncClosed} through the Hamilton's principle:
\begin{subequations}\label{ELPart}
\begin{align}
m_a\ddot q_a+\der_{q_a}V-\sum_{j=1}^{(\mathcal{N}-2)/2}\frac{\gamma_a^j}{\omega_a^j}(Q_a^j-\frac{\gamma_a^j}{(\omega_a^j)^2}q_a)&=0,\label{ELPart:a}\\
M\ddot Q_a^j+\omega_a^j(Q_a^j-\frac{\gamma_a^j}{(\omega_a^j)^2}q_a)&=0.\label{ELPart:b}
\end{align}
\end{subequations}
We can eliminate the degrees of freedom of the heat bath $Q_a^j$ from \eqref{ELPart:b} using analogous arguments as in the Hamiltonian side. This process leads from 
\eqref{ELPart:a} to
\begin{equation}
m_a\ddot{q}_a+\der_{q_a}V + \int _0^t dt^{\prime} \gamma_a(t - t^{\prime})\dot{q}_a = F(t),
\label{Langevin2}
\end{equation}
which is the Langevin equation for a  Brownian motion, i.e.~the random motion of a particle immersed in a fluid and being subject to dissipation and fluctuation \cite{Zwanzig}.

By assuming that the environment is Markovian  
\begin{equation}
m_a\ddot{q}_a+\der_{q_a}V +  \gamma_a\dot{q}_a = F(t),
\label{Langevin3}
\end{equation}
which accounts for the Lagrangian Continuous Damping Reduction \eqref{LagDR}. Furthermore, it is easy to see that \eqref{Langevin3} can be equivalently obtained from the Lagrange-d'Alembert principle \eqref{LdAPrinL} using $L^s(q_a,\dot q_a)$ \eqref{LagSHB:a} and external forces $f_L(q_a,\dot q_a)$ \eqref{Forces:L} given by the damped matrix $\tilde\Lambda$ \eqref{DampedMatrix}. Finally, after aplying the Legendre transform given by the Lagrangian of the open subsystem, i.e. $\F L^s$, over the Lagrangian itself and the external forces, we can employ the Lagrange-d'Alembert principle in its Hamiltonian form \eqref{LdAPrinH} with $H^s(q_a,p_a)$ \eqref{HamSHB:a} and $f_H(q_a,p_a)$ \eqref{Forces:H} (again $\tilde\Lambda$ in \eqref{DampedMatrix}) in order to reobtain \eqref{HamOpenFinal}. This sets the commutativity of diagram \eqref{DiagramCont} for the heat bath, as established in Theorem \ref{ContTheo}.

Note finally that we have set $n=2$ for simplicity in the figures, but we have not made any further restriction on the index $a$ along the example. Therefore, this procedure applies  to any $n<\mathcal{N}.$

\section{Discrete counterpart}\label{DiscCount}

 According to \S\ref{DiscreteFramework} we set the following discretisation $L_d(\mathfrak{q}_k,\mathfrak{q}_{k+1}):(\R^d\times\R^d)^N\Flder\R$ of the Lagrangian \eqref{LagN}, where we define $(\mathfrak{q}_k,\mathfrak{q}_{k+1}):=((q_1^k, q_1^{k+1}),(q_2^k, q_2^{k+1}),...,(q_{\mathcal{N}}^k,q_{\mathcal{N}}^{k+1}))$\footnote{Henceforth, we will rise the discrete time index $k$ when convenient in order to clarify the notation.} and $\tilde{\mathfrak{q}}_{k+1}:=(q_1^{k+1},...,q_{n}^{k+1})$ (the same for  $k$):
\begin{equation}\label{DiscLagN}
\begin{split}
L_d(\mathfrak{q}_k,\mathfrak{q}_{k+1},\gamma)&=\sum_{i=1}^{\mathcal{N}}\frac{1}{2h}m_{i} (q_i^{k+1}-q_i^{k})^2-h\,V(\gamma\tilde{\mathfrak{q}}_{k}+(1-\gamma)\tilde{\mathfrak{q}}_{k+1})\\
&\quad\quad\quad-\frac{h}{2}\sum_{i,j=1}^{\mathcal{N}}\,\lambda_{ij}(\gamma q_i^{k}+ (1-\gamma)q_i^{k+1})(\gamma q_j^{k}+ (1-\gamma)q_j^{k+1}),
\end{split}
\end{equation}
with $\gamma\in[0,1]$. The discrete Euler-Lagrange equations \eqref{DEL} in this case read
\begin{equation}\label{DiscELEqsN}
\begin{split}
&m_i\frac{q_i^{k+1}-2q_i^{k}+q_i^{k-1}}{h^2}+\gamma\der_{q_i} V(\gamma\tilde{\mathfrak{q}}_{k}+(1-\gamma)\tilde{\mathfrak{q}}_{k+1})+(1-\gamma)\der_{q_i} V(\gamma\tilde{\mathfrak{q}}_{k-1}+(1-\gamma)\tilde{\mathfrak{q}}_{k})
\\
&\hspace{2.5cm} +\gamma\sum_{j=1}^{\mathcal{N}}\lambda_{ij}(\gamma q_j^{k}+(1-\gamma)q_j^{k+1})+(1-\gamma)\sum_{j=1}^{\mathcal{N}}\lambda_{ij}(\gamma q_j^{k-1}+(1-\gamma)q_j^{k})=0,
\end{split}
\end{equation}
for $i=1,...,\mathcal{N}$ and $k=1,...,N-1$. We will denote these equations DELEqs$_{\tiny\mathcal{N}}^{\gamma}$  in the diagram below (after Discrete Euler Lagrange Equations), accounting for the closed system containing the $\mathcal{N}$ particles and depending on the $\gamma$ parameter. As discussed above, \eqref{DiscELEqsN} are a discretisation in finite differences of \eqref{ELEqsN}.

\begin{remark}
{\rm Observe that both $\gamma=1$ and $\gamma=0$, corresponding to the initial and final endpoint discretisations, respectively, provide the same discrete EL equations:
\[
\gamma=0,1:\quad\quad\quad m_i\frac{q_i^{k+1}-2q_i^{k}+q_i^{k-1}}{h^2}+\der_{q_i} V(\tilde{\mathfrak{q}}_{k})+\sum_{j=1}^{\mathcal{N}}\lambda_{ij}q_j^{k}=0,
\]
whereas $\gamma=1/2$ (midpoint rule) leads to
\[
\begin{split}
&\gamma=\frac{1}{2}:\quad m_i\frac{q_i^{k+1}-2q_i^{k}+q_i^{k-1}}{h^2}+\frac{1}{2}\der_{q_i} V\Big(\frac{\tilde{\mathfrak{q}}_{k}+\tilde{\mathfrak{q}}_{k+1}}{2}\Big)+
\frac{1}{2}\der_{q_i} V\Big(\frac{\tilde{\mathfrak{q}}_{k-1}+\tilde{\mathfrak{q}}_{k}}{2}\Big)\\
&\hspace{6cm} +\frac{1}{2}\sum_{j=1}^{\mathcal{N}}\lambda_{ij}\frac{q_j^{k}+q_j^{k+1}}{2}+\frac{1}{2}\sum_{j=1}^{\mathcal{N}}\lambda_{ij}\frac{q_j^{k-1}+q_j^{k}}{2}=0.
\end{split}
\]} \hfill $\diamond$
\end{remark}
The discrete Legendre transforms \eqref{DLT} over \eqref{DiscLagN} read 
\[
\begin{split}
p_i^k&=m_i\frac{q_i^{k+1}-q_i^k}{h}\,\,\,+\,\,\,\,\,\,h\gamma\, \der_{q_i} V(\gamma\tilde{\mathfrak{q}}_{k}+ (1-\gamma)\tilde{\mathfrak{q}}_{k+1})\,\,\,+\,\,\,\,\,h\gamma\sum_{j=1}^{\mathcal{N}}\lambda_{ij}(\gamma q_j^k+(1-\gamma)q_j^{k+1}),\\
p_i^{k+1}&=m_i\frac{q_i^{k+1}-q_i^k}{h}-h(1-\gamma) \der_{q_i} V(\gamma\tilde{\mathfrak{q}}_{k}+ (1-\gamma)\tilde{\mathfrak{q}}_{k+1})-h(1-\gamma)\sum_{j=1}^{\mathcal{N}}\lambda_{ij}(\gamma q_j^k+(1-\gamma)q_j^{k+1}).
\end{split}
\]
From these equations it is easy to check that
\[
(1-\gamma)p_i^k+\gamma p^{k+1}_i=m_i\frac{q_i^{k+1}-q_i^k}{h}.
\]
This equation + momentum mathching \eqref{MomMat} provides the symplectic integrator:
\begin{subequations}\label{DHEqsClosed}
\begin{align}
q_i^{k+1}&=q_i^k+h\frac{(1-\gamma)p_i^k+\gamma p^{k+1}_i}{m_i},\label{DHEqsClosed:a}\\
p_i^{k+1}&=p_i^k-h\der_{q_i} V(\gamma\tilde{\mathfrak{q}}_{k}+ (1-\gamma)\tilde{\mathfrak{q}}_{k+1})-h\sum_{j=1}^{\mathcal{N}}\lambda_{ij}(\gamma q_j^k+(1-\gamma)q_j^{k+1}),\label{DHEqsClosed_b}
\end{align}
\end{subequations}
which is implicit in both $q$ and $p$. Similarly  to the Lagrangian side, we will denote these equations in the diagram below as DHEqs$_{\tiny\mathcal{N}}^{\gamma}$ (after Discrete Hamilton Equations).

\begin{remark}\label{SemiImpRem}
{\rm Note that $\gamma=0$ and $\gamma=1$ generate, respectively,
\[
\begin{split}
\gamma=0:\quad\quad q_i^{k+1}&=q_i^k+h\frac{p_i^k}{m_i},\quad\,\,\,\,\,\,p_i^{k+1}=p_i^k-h\der_{q_i} V(\tilde{\mathfrak{q}}_{k+1})-h\sum_{j=1}^{\mathcal{N}}\lambda_{ij}q_j^{k+1},\\
\gamma=1:\quad\quad q_i^{k+1}&=q_i^k+h\frac{p_i^{k+1}}{m_i},\quad\,\,p_i^{k+1}=p_i^k\,-\,h\der_{q_i} V(\tilde{\mathfrak{q}}_{k})\,\,\,-\,\,\,h\sum_{j=1}^{\mathcal{N}}\lambda_{ij}q_j^{k}.
\end{split}
\]
for $i=1,...,\mathcal{N}$ and $k=1,...,N-1;$ which are both symplectic Euler \cite{HLW, SS} discretisations of \eqref{HamEqsN}. Note the semi-implicit nature of both integrators, i.e. implicit in $q$ and explicit in $p$ for $\gamma=0$ and conversely for $\gamma=1$ Finally, $\gamma=1/2$ generates naturally the midpoint rule
\[
\gamma=\frac{1}{2}:\quad\,\, q_i^{k+1}=q_i^k+h\frac{p_i^k+p_i^{k+1}}{2m_i},\quad p_i^{k+1}=p_i^k-h\der_{q_i} V\Big(\frac{\tilde{\mathfrak{q}}_{k}+\tilde{\mathfrak{q}}_{k+1}}{2}\Big)-h\sum_{j=1}^{\mathcal{N}}\lambda_{ij}\Big( \frac{q_j^{k}+q_j^{k+1}}{2}\Big),
\]
again implicit in both $q$ and $p$.
} \hfill $\diamond$
\end{remark}

\subsection{Discrete Damping Reduction} 

The discrete version of the Continuous Damping Reduction \eqref{LagDR} shall be called Discrete Damping Reduction (DDR henceforth). Before its definition, let us consider the discretisation of the Lagrangian open subsystem \eqref{Subsystem:L}, which consistently with \eqref{DiscLagN} is given by
\begin{equation}\label{DLagOpen}
L^s_d(\tilde{\mathfrak{q}}_k,\tilde{\mathfrak{q}}_{k+1},\gamma)=\sum_{a=1}^n\frac{1}{2h}m_a (q_a^{k+1}-q_a^k)^2-hV(\gamma\tilde{\mathfrak{q}}_{k}+ (1-\gamma)\tilde{\mathfrak{q}}_{k+1}),
\end{equation}
plus the discrete external forces
\begin{equation}\label{DiscExtForces}
f^{-}_{L_d^s}(\tilde{\mathfrak{q}}_k,\tilde{\mathfrak{q}}_{k+1},\gamma)=\gamma\sum_{b=1}^{n}\tilde\lambda_{ab}(q_b^{k+1}-q_b^{k}),\quad f^{+}_{L_d^s}(\tilde{\mathfrak{q}}_k,\tilde{\mathfrak{q}}_{k+1},\gamma)_a=(1-\gamma)\sum_{b=1}^{n}\tilde\lambda_{ab}(q_b^{k+1}-q_b^{k}).
\end{equation}
The forced discrete Legendre transform \eqref{DForcedLT} for this system reads
\begin{subequations}\label{FLTFinal}
\begin{align}
p_a^k&=m_a\frac{q_a^{k+1}-q_a^k}{h}+\,\,\,\,h\,\gamma\,\,\der_{q_a}V(\gamma\tilde{\mathfrak{q}}_{k}+ (1-\gamma)\tilde{\mathfrak{q}}_{k+1})\,\,\,-\,\,\,\gamma\sum_{b=1}^n\tilde\lambda_{ab}(q_a^{k+1}-q_a^k),\label{FLTFinal:a}\\
p_a^{k+1}&=m_a\frac{q_a^{k+1}-q_a^k}{h}+h\,(1-\gamma)\der_{q_a}V(\gamma\tilde{\mathfrak{q}}_{k}+ (1-\gamma)\tilde{\mathfrak{q}}_{k+1})+(1-\gamma)\sum_{b=1}^n\tilde\lambda_{ab}(q_a^{k+1}-q_a^k),\label{FLTFinal:b}
\end{align}
\end{subequations}
which furthermore yields
\begin{equation}\label{MomentaRelation}
(1-\gamma)p_a^k+\gamma p_a^{k+1}=m_a\frac{q_a^{k+1}-q_a^k}{h}.
\end{equation}
\begin{definition}\label{DefDDR}
 For $i,j=1,...,\mathcal{N};\, \,a,b=1,...,n; \,k=1,...,N-1$ and $\gamma\in[0,1]$ we define the {\rm DDR} in its Lagrangian and Hamiltonina forms, respectively, as
\begin{subequations}\label{DDR}
\begin{align}
\sum_{j=1}^{\mathcal{N}}\lambda_{ij}(\gamma q_j^k+(1-\gamma)q_j^{k+1})&\rightarrow -\sum_{b=1}^n\tilde\lambda_{ab}\frac{q_b^{k+1}-q_b^{k}}{h},\label{DDR:L}\\
\sum_{j=1}^{\mathcal{N}}\lambda_{ij}(\gamma q_j^k+(1-\gamma)q_j^{k+1})&\rightarrow -\sum_{b=1}^n\tilde\lambda_{ab}\frac{(1-\gamma)p_b^k+\gamma p_b^{k+1}}{m_b}\label{DDR:H}
\end{align}
\end{subequations}
where \eqref{DDR:H} follows directly from \eqref{DDR:L} and \eqref{MomentaRelation}.
\end{definition}
With these elements we can establish the following theorem.

\begin{theorem}\label{DiscTheo}
Given a discrete Lagrangian function $L_d(\mathfrak{q}_k,\mathfrak{q}_{k+1},\gamma)$ \eqref{DiscLagN} for a closed system of $\mathcal{N}$ interacting particles,  the discrete Lagrangian of the open subsystem $L_d^s(\tilde{\mathfrak{q}}_k,\tilde{\mathfrak{q}}_{k+1},\gamma)$ \eqref{DLagOpen}, the {\rm DDR} given in Definition \ref{DefDDR} and the discrete forces \eqref{DiscExtForces}, the following diagram commutes. 

{\rm \begin{equation}\label{DiscDiag}
\xymatrix{
*[F]{\begin{array}{c}
\\
L_d(\mathfrak{q}_k,\mathfrak{q}_{k+1},\gamma),\,\eqref{DiscLagN}
\\\\
\end{array}}\ar[r]^{\F^{\pm} L_d}\ar[d]_{\mbox{DHP} \,\eqref{DEL}}  &  *[F]{\begin{array}{c}
\\
\mbox{DHEqs}_{\tiny\mathcal{N}}^{\gamma},\,\,\eqref{DHEqsClosed}
\\\\
\end{array}}\ar[ddd]^{\mbox{DDR}\,\eqref{DDR:H}} \\
*[F]{\begin{array}{c}
\\
\mbox{DELEqs}_{\tiny\mathcal{N}}^{\gamma},\,\,\eqref{DiscELEqsN}
\\\\
\end{array}}\ar[d]_{\mbox{DDR}\,\eqref{DDR:L}} & \\
*[F]{\begin{array}{c}
\\
\mbox{FDELEqs}_n^{\gamma},\,\eqref{ForcedEqs:n}
\\\\
\end{array}}\ar@{<=>}[d]_{\mbox{DLdA}\,\eqref{ForcedDEL}} &  \\
*[F]{\begin{array}{c}
\\
L_d^s(\tilde{\mathfrak{q}}_k,\tilde{\mathfrak{q}}_{k+1},\gamma)\,\eqref{DLagOpen}
\\
f_{L_d^s}^{-}(\tilde{\mathfrak{q}}_k,\tilde{\mathfrak{q}}_{k+1},\gamma),\,\,\,f_{L_d^s}^{+}(\tilde{\mathfrak{q}}_{k},\tilde{\mathfrak{q}}_{k+1},\gamma)\,\,\,\, \eqref{DiscExtForces}\\
 \\
\end{array}}\ar@/_4pc/[r]_{\F^{\pm} (L_d^s)^{\tiny \mbox{f}}} & *[F]{\begin{array}{c}
\\
\mbox{FDHEqs}_n^{\gamma}
\\\\
\end{array}}
}
\end{equation}}
In other words, its two branches, i.e.
{\rm\begin{itemize}
\item[1.] $\F^{\pm}(L_d^s)^{\tiny\mbox{f}}$ $\,\circ$ DLdA $\circ$ DDR(Lag) $\circ$ DHP

\item[2.] DDR(Ham) $\circ$ $\F^{\pm} L_d$,
\end{itemize}}
\hspace{-0.55cm} provide  the same set of Forced Discrete Hamilton Equations ({\rm FDHEqs$_n^{\gamma}$} ) of the subsystem of $n$ damped particles.
\end{theorem}

\begin{proof}
Applying \eqref{DDR:H} to the DHEqs$_{\tiny\mathcal{N}}^{\gamma}$ \eqref{DHEqsClosed} we obtain
\begin{subequations}\label{DHEqsOpen}
\begin{align}
q_a^{k+1}&=q_a^k+h\frac{(1-\gamma)p_a^k+\gamma p^{k+1}_a}{m_a},\label{DHEqsOpen:a}\\
p_a^{k+1}&=p_a^k-h\der_{q_a} V(\gamma\tilde{\mathfrak{q}}_{k}+ (1-\gamma)\tilde{\mathfrak{q}}_{k+1})+h\sum_{b=1}^n\tilde\lambda_{ab}\frac{(1-\gamma)p_a^k+\gamma p_a^{k+1}}{m_b},\label{DHEqsOpen_b}
\end{align}
\end{subequations}
which are $\mbox{FDHEqs}_n^{\gamma}$. This accounts for branch 2 in the diagram.

On the other hand, we observe that the application of the Lagrangian DDR \eqref{DDR:L} over \eqref{DiscELEqsN} produces
\begin{equation}\label{ForcedEqs:n}
\begin{split}
&m_a\frac{q_a^{k+1}-2q_a^{k}+q_a^{k-1}}{h^2}+\gamma\der_{q_a} V(\gamma\tilde{\mathfrak{q}}_{k}+(1-\gamma)\tilde{\mathfrak{q}}_{k+1})
\\
& \,\,\,\,\,\,\,\,+(1-\gamma)\der_{q_a} V(\gamma\tilde{\mathfrak{q}}_{k-1}+(1-\gamma)\tilde{\mathfrak{q}}_{k})= \gamma\sum_{b=1}^{n}\tilde\lambda_{ab}\frac{q_b^{k+1}-q_b^k}{h}+(1-\gamma)\sum_{b=1}^{n}\tilde\lambda_{ab}\frac{q_b^{k}-q_b^{k-1}}{h},
\end{split}
\end{equation}
which are denoted FDELEqs$_{n}^{\gamma}$   (after Forced Discrete Euler Lagrange Equations) in the diagram, accounting for the open system containing the $n$ particles and depending on the $\gamma$ parameter. It is easy to see that the choice of \eqref{DLagOpen} for the discrete Lagrangian of the open subsystem and \eqref{DiscExtForces} for the discrete forces, also provides $\mbox{FDELEqs}_n^{\gamma}$ through the Lagrange-d'Alembert principle \eqref{ForcedDEL}. Furthermore, \eqref{ForcedEqs:n}  are equivalent to the momentum mathching condition $(p_a^k)^{-}=(p_a^k)^{+}$ for \eqref{FLTFinal:a}, \eqref{FLTFinal:b}. Using \eqref{MomentaRelation} and \eqref{FLTFinal}, one arrives to
\[
\begin{split}
q_a^{k+1}&=q_a^k+h\frac{(1-\gamma)p_a^k+\gamma p^{k+1}_a}{m_a},\\
p_a^{k+1}&=p_a^k-h\der_{q_a} V(\gamma\tilde{\mathfrak{q}}_{k}+ (1-\gamma)\tilde{\mathfrak{q}}_{k+1})+\sum_{b=1}^n\tilde\lambda_{ab}(q_b^{k+1}-q_k^b),
\end{split}
\]
which, after considering that
\[
q_a^{k+1}-q_a^k=h\frac{(1-\gamma)p_a^k+\gamma p_a^{k+1}}{m_a}
\]
that follows directly from \eqref{MomentaRelation}, are again $\mbox{FDHEqs}_n^{\gamma}$ and accounts for branch 1 in the diagram, proving its commutativity.
\end{proof}

\begin{remark}\label{ForcedSemiImpliRemark}
{\rm
Naturally, $\mbox{FDHEqs}_n^{\gamma}$ are not symplectic integrators anymore, since they are approximating the dynamics of non-Hamiltonian systems. However, we observe that $\mbox{FDHEqs}_n^{0}$ and $\mbox{FDHEqs}_n^{1}$, i.e.
\[
\begin{split}
\gamma=0:\quad\quad\quad q_a^{k+1}&=q_a^k+h\frac{p_a^k}{m_a},\quad\,\,\,\,p_a^{k+1}=p_a^k-h\der_{q_a} V(\tilde{\mathfrak{q}}_{k+1})+h\sum_{b=1}^n\tilde\lambda_{ab}\frac{p_b^k}{m_b},\\
\gamma=1:\quad\quad\quad q_a^{k+1}&=q_a^k+h\frac{p_a^{k+1}}{m_a},\quad\,\,p_a^{k+1}=p_a^k-h\der_{q_a} V(\tilde{\mathfrak{q}}_{k})+h\sum_{b=1}^n\tilde\lambda_{ab}\frac{p_b^{k+1}}{m_b},
\end{split}
\]
retain the semi-implicit nature of their symplectic counterparts for the closed system shown in Remark \ref{SemiImpRem}. Finally $\mbox{FDHEqs}_n^{1/2}$
\[
\gamma=\frac{1}{2}:\quad q_a^{k+1}=q_a^k+h\frac{p_a^k+p^{k+1}_a}{2m_a},\,\,\, p_a^{k+1}=p_a^k-h\der_{q_a} V\Big(\frac{\tilde{\mathfrak{q}}_{k}+\tilde{\mathfrak{q}}_{k+1}}{2}\Big)+h\sum_{b=1}^{n}\tilde\lambda_{ab}\Big( \frac{p_b^{k}+p_b^{k+1}}{2m_b}\Big),
\]
is fully implicit as its symplectic counterpart.
}\hfill $\diamond$
\end{remark}

\section{Rubin Model (Two Masses)}\label{RubinSec}

\subsection{Continuous setting}

\begin{figure}[!htb]
\includegraphics[scale=0.28]{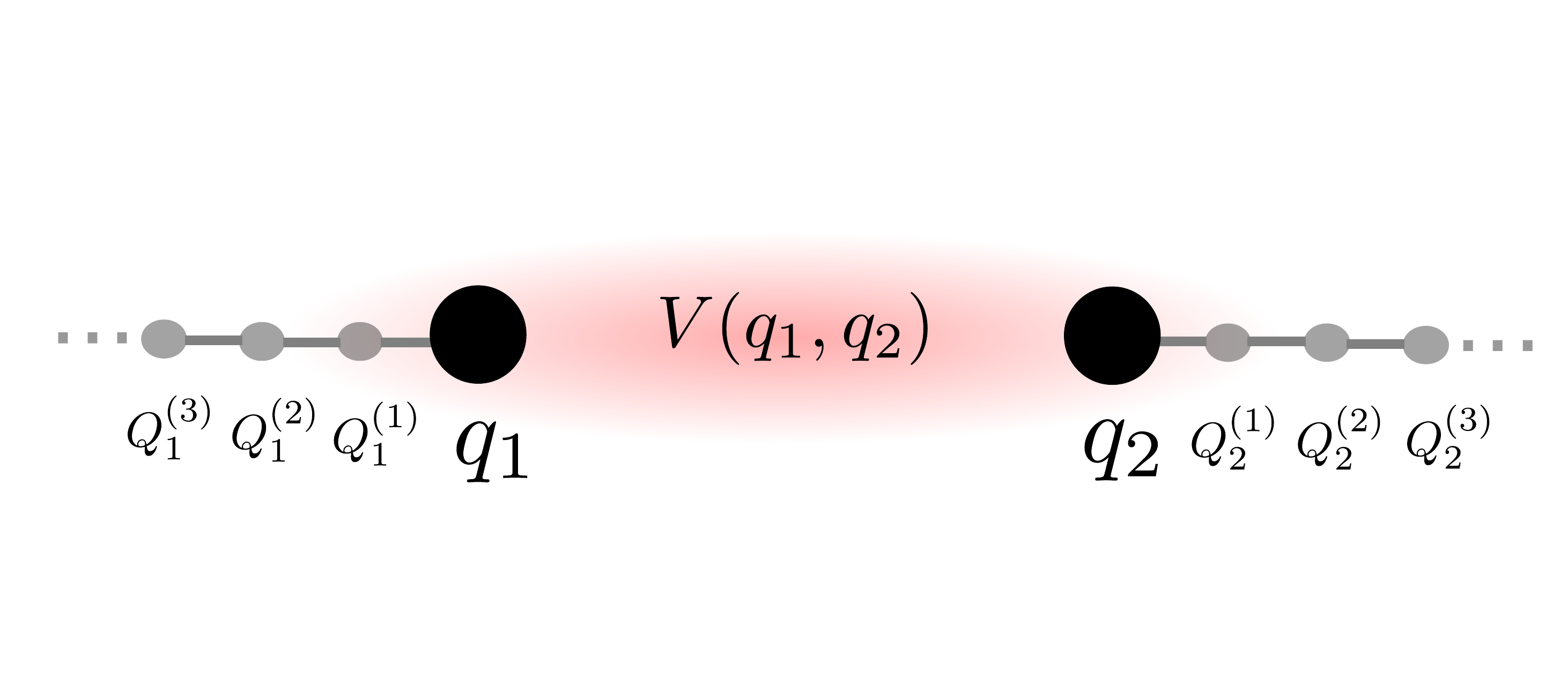}
\caption{Two systems dissipated by transmission lines interacting through an arbitrary interaction potential.}
\label{fig2}
\end{figure}

A different approach from the heat bath \S\ref{HBSection} in modeling the linear dissipative environment is to represent it in terms of a string of systems coupled  to the main (open) system in a linear fashion (\cite{Yurke}) as illustrated in Figure \ref{fig2}. In this representation, the string (which we will call {\it transmission line}) transmits the energy of the open system down the line in terms of outward traveling waves, and thus yields dissipative effects. 

As in \S\ref{HBSection} we assume that there are $(\mathcal{N}-2)/2$ in each transmission line; moreover, we also employ the notation \eqref{Notation}. The Lagrangian of the closed system illustrated in Figure \ref{fig2} reads as following:
\[
L(\mathfrak{q},\dot{\mathfrak{q}})= \sum_{a=1}^2 L^s(q_a, \dot q_a) + \sum_{a=1}^2 \sum_{j=1}^{(\mathcal{N}-2)/2} L^{\tiny\mbox{tl}}(Q_a^j, \dot Q_a^j) + L^{\tiny\mbox{int}}(q_1,q_2,Q_1^1,Q_2^1),
\]
where $L^s$ encodes the dynamics of the open systems $a =1,2$, $L^{\tiny\mbox{tl}}$ encodes the dynamics of their corresponding environment (transmission lines) $j = 1,2,...,(\mathcal{N}-2)/2$, and $L^{\tiny\mbox{int}}$ encodes the system-environment interaction. In particular:
\begin{subequations}\label{LagLines}
\begin{align}
L^s(q_a, \dot{q}_a) &= \,\,\,\,\frac{1}{2}m_a \dot{q}_a^2 - V(q_1,q_2), \label{LagLines:a}\\
L^{\tiny\mbox{tl}}(Q_a^j,\dot Q_a^j) &=\,\,\,\,\frac{1}{2}m_a^j (\dot Q_a^j)^2 - \frac{1}{2}\lambda_a(Q_a^{j+1} - Q_a^j)^2, \\
L^{\tiny\mbox{int}}(q_1,q_2,Q_1^1,Q_2^1) &= - \frac{1}{2}\lambda_1 (Q_1^1 - q_1)^2- \frac{1}{2}\lambda_2 (Q_2^1 - q_2)^2. \label{LagLines:c}
\end{align} 
\end{subequations}
Here we will assume that $m_a^j = M$ for all $a$ and $j$.  
Moreover, each transmission line is coupled to the corresponding open system $q_a$ through the same coupling strength with which its elements are coupled together, namely $\lambda_a$. We observe that Figure \ref{fig2} and \eqref{LagLines} correspond to the following quadratic interaction matrix $\Lambda$, which we show decomposed in boxes for simplicity: 
\begin{small}
\[
2\Lambda_0=\begin{bmatrix}
\lambda_1 & 0 &-\lambda_1 & 0\\
 0 &\lambda_2 & 0 & -\lambda_2\\
-\lambda_1 &0 &\lambda_1 & 0\\
0&-\lambda_2 &0&\lambda_2
\end{bmatrix}, \,\,2\Lambda_1=\begin{bmatrix}
 \lambda_1& -\lambda_1 &0& \cdots&\cdots  &0\\
-\lambda_1& \lambda_1 & -\lambda_1 & 0& \cdots& 0\\
 0&-\lambda_1&\lambda_1&-\lambda_1&\cdots& 0\\
\vdots &\vdots&\ddots&\ddots&\ddots&\vdots\\
 0&\cdots &0 &-\lambda_1 & \lambda_1&-\lambda_1\\
0 &0&\cdots&0&-\lambda_1& \lambda_1
\end{bmatrix}.
\]
\end{small}
$\Lambda_0$ encodes the quadratic interaction among the particles $\lc q_1,q_2,Q_1^1,Q_2^1\rc$ (in this precise order both in rows and columns), whereas $\Lambda_1$ represents the same for the set $\lc Q_1^1,Q_1^2,\cdots,Q_1^{(\mathcal{N}-2)/2}\rc$ (in order to clarify the notation, we remark that the $\delta\,$-th $\eta$-th entry in $\Lambda_1$ would be the $(\delta+1)$-th $(\gamma+1)$-th in $\Lambda$, since the latter incorporates rows and columns for $q_1,q_2$, whereas the ones corresponding to $Q_2^1$ are displaced from $\Lambda_0$ to $\Lambda_2$ as we shall comment shortly). We observe that $\Lambda_1$ is 0 except for the diagonal and upper and lower diagonal entries, which physically means that each particle in the transmission line only interacts with its first neighbours (and its own quadratic potential $\lambda_1(Q_1^j)^2/2$). The interaction among $\lc Q_2^1,Q_2^2,\cdots,Q_2^{(\mathcal{N}-2)/2}\rc$ is given by $\Lambda_2$, which is equivalent to $\Lambda_1$ with $1\Flder 2$ (according to the notation above, the $\delta\,$-th $\eta$-th entry in $\Lambda_2$ corresponds to $(\delta+\frac{\mathcal{N}-4}{2})$-th $(\eta+\frac{\mathcal{N}-4}{2})$-th in $\Lambda$).

The Euler-Lagrange equations for the closed system \eqref{LagLines} read as following:
\begin{subequations}\label{conWave}
\begin{align}
m_a\ddot{q}_a+ \der_{q_a}V - \lambda_a (Q_a^1 - q_a) &= 0, \quad a= 1,2; \label{conWave01}\\ 
M\ddot{Q}_a^j - \lambda_a ( Q_a^{j+1}-2Q_a^j+Q_a^{j-1})&= 0,\quad  a = 1,2;\,\, j =2,...,(\mathcal{N}-2)/2.
\label{conWave02}
\end{align}
\end{subequations}
As the number of particles forming the transmission line tends to infinity, i.e. $\mathcal{N}\Flder\infty$, while the length of the line remains finite, the equations \eqref{conWave} tend to a set of spatially continuous equations:
\begin{subequations}\label{WaveReduced}
\begin{align}
m_a\ddot{q}_a+ \der_{q_a}V - K_a \frac{\partial q_a}{\partial x} &= 0,\quad a = 1,2;\label{conWave03} \\
\rho\, \ddot{q}_a - K_a \frac{\partial ^2 q_a}{\partial x^2}&= 0, \quad a = 1,2;
\label{conWave04}
\end{align}
\end{subequations}
where $\rho = M/h_x$ is the mass density of the transmission line ($h_x$ being the spatial separation of the systems in the line), and $K_a = h_x \lambda_a$. As it is obvious from \eqref{WaveReduced}, we are considering $q_a$ as a function of space and time, i.e. $q_a=q_a(t,x)$ (we consider that the transmission line is distributed along the $x$ direction of $\R^d$). \eqref{conWave04} is a wave equation which admits the following solution in terms of inward and outward traveling waves with the velocity $v_a = \sqrt{K_a/\rho}$ :
\[
q_a(t,x) = q_a^{\tiny\mbox{in}}(x+v_at) + q_a^{\tiny\mbox{out}}(x-v_at). 
\]
Now, we will make a crucial assumption, $q_a^{\tiny\mbox{in}}(x+v_at) = 0$; that is, the transmission line is long enough compared to the time scale of system dynamics, and thus, there will not be any waves traveling towards the open system as a result of reflection.\footnote{We refer to \cite{DavidsPaper} for details on the justification of this assumption.} With this assumption it is easy to check the following identity 
\begin{equation}
\frac{\partial q_a(t,x)}{\partial x} = -\frac{1}{v_a}\frac{\partial q_a(t,x)}{\partial t}.
\label{SpaTemp}
\end{equation}
In other words, the wave solution for the behavior of the transmission line converts the spatial derivatives into temporal derivatives. Consequently, \eqref{conWave03} can be written in the following dissipative form
\begin{equation}
m_a\ddot{q}_a+ \der_{q_a}V + D_a  \dot{q}_a = 0, \quad a = 1,2;
\label{conDissLag}
\end{equation} 
with $D_a = \sqrt{K_a\rho}$ being the damping rate. Thus, by eliminating the environment (transmission line, whose dynamics is given by \eqref{conWave02}), from the closed system dynamics (both equations \eqref{conWave}), the resulting dynamics for the two-mass system (open system) appears dissipative. This represents the Lagrangian CDR for the system given in Figure \ref{fig2}, where \eqref{conDissLag} accounts for \eqref{LagAfterDR}. The damping matrix $\tilde\Lambda$ reads
\begin{equation}\label{tildeLambdaLines}
\tilde\Lambda=\begin{bmatrix}
D_1&0\\
0&D_2
\end{bmatrix}.
\end{equation}

From \eqref{LagLines}, using the Legendre transformation one obtains the Hamiltonian function
\[
H(\mathfrak{q},\mathfrak{p})= \sum_{a=1}^2 H^s(q_a, p_a) + \sum_{a=1}^2 \sum_{j=1}^{(\mathcal{N}-2)/2} H^{\tiny\mbox{tl}}(Q_a^j, P_a^j) - L^{\tiny\mbox{int}}(q_1,q_2,Q_1^1,Q_2^1),
\]
where
\begin{eqnarray}
H^s(q_a, \dot{q}_a) &=& \,\,\,\,\frac{1}{2\,m_a} p_a^2 + V(q_1,q_2), \label{HamSubsystems}\\
H^{\tiny\mbox{tl}}(Q_a^j,\dot Q_a^j) &=&\,\,\,\,\frac{1}{2\,M} (P_a^j)^2 + \frac{1}{2}\lambda_a(Q_a^{j+1} - Q_a^j)^2, \nonumber
\end{eqnarray}
and $L^{\tiny\mbox{int}}$ is given in \eqref{LagLines:c}. Using the same arguments which led to \eqref{WaveReduced}, we observe that the dynamics of the open system can be written in terms of the Hamilton's equations as following:
\[
\dot{q}_a = \frac{1}{m_a}p_a,\quad
\dot{p}_a =-\der_{q_a}V + K_a \frac{\partial q_a}{\partial x}.
\]
Employing the relationiship \eqref{SpaTemp}, these equations can be re-written in the following form:
\[
\dot{q}_a= \frac{1}{m_a}p_a,\quad
\dot{p}_a = -\partial_{q_a} V - \frac{D_a}{m_a} p_a.
\label{conHam02}
\]
These equations represent the dynamics of the forced (open) system after the Hamiltonian CDR. With these equations and the particular expressions of the subsystem Lagrangian $L^s$ \eqref{LagLines:a}, the subsystem Hamiltonian \eqref{HamSubsystems} and the damping matrix \eqref{tildeLambdaLines}, it is easy to establish the commutativity prescribed in Theorem \ref{ContTheo}.

\begin{remark}
{\rm The generalisation of the previous development to several transmission lines attached to each particle in the open subsystem is straightforward. We observe that each transmission line must be placed along  one of the independent spacial directions in $\R^d$, such that there are no wave projections over the other ones (we sketch such a system in Figure \ref{SeveralLines}, where naturally the projection is inevitable).  $Q_{(a,\alpha)}^j$ would represent the $j$-th particle of the $\alpha$-th transmission line attached to the $a$-th particle in the open subsystem, all of them with mass $M$. Assuming that there are a total of $\mathcal{N}$ particles, and $d$ (at most) transmission lines for each $a$, then $a=1,2$; $\alpha=1,...,d$ and $j=1,...,(\mathcal{N}-2)/2d$. Furthermore, we assume that there are only outgoing waves in each direction, i.e. {\rm $q_a(t,x_{\alpha})=q^{\tiny\mbox{out}}_a(x_{\alpha}-v_{(a,\alpha)}t)$}, where $v_{(a,\alpha)}=\sqrt{K_{(a,\alpha)}/\rho_{\alpha}}$, $K_{(a,\alpha)}=h_{x_{\alpha}}\lambda_a$ and $\rho_{\alpha}=M/h_{x_{\alpha}}$; with $h_{x_{\alpha}}$ the spatial separation of the system in the $x_{\alpha}$-line. Taking into account  the superposition of waves, following the showed CDR we end up with the damped equation 
\[
m_a\ddot{q}_a+ \der_{q_a}V + \tilde D_a  \dot{q}_a = 0, \quad a = 1,2;
\]
with $\tilde D_a=\sum_{\alpha=1}^dD_{(a,\alpha)}$ and $D_{(a,\alpha)}=\sqrt{K_{(a,\alpha)}\rho_{\alpha}}$.
\begin{figure}[!htb]
\includegraphics[scale=0.28]{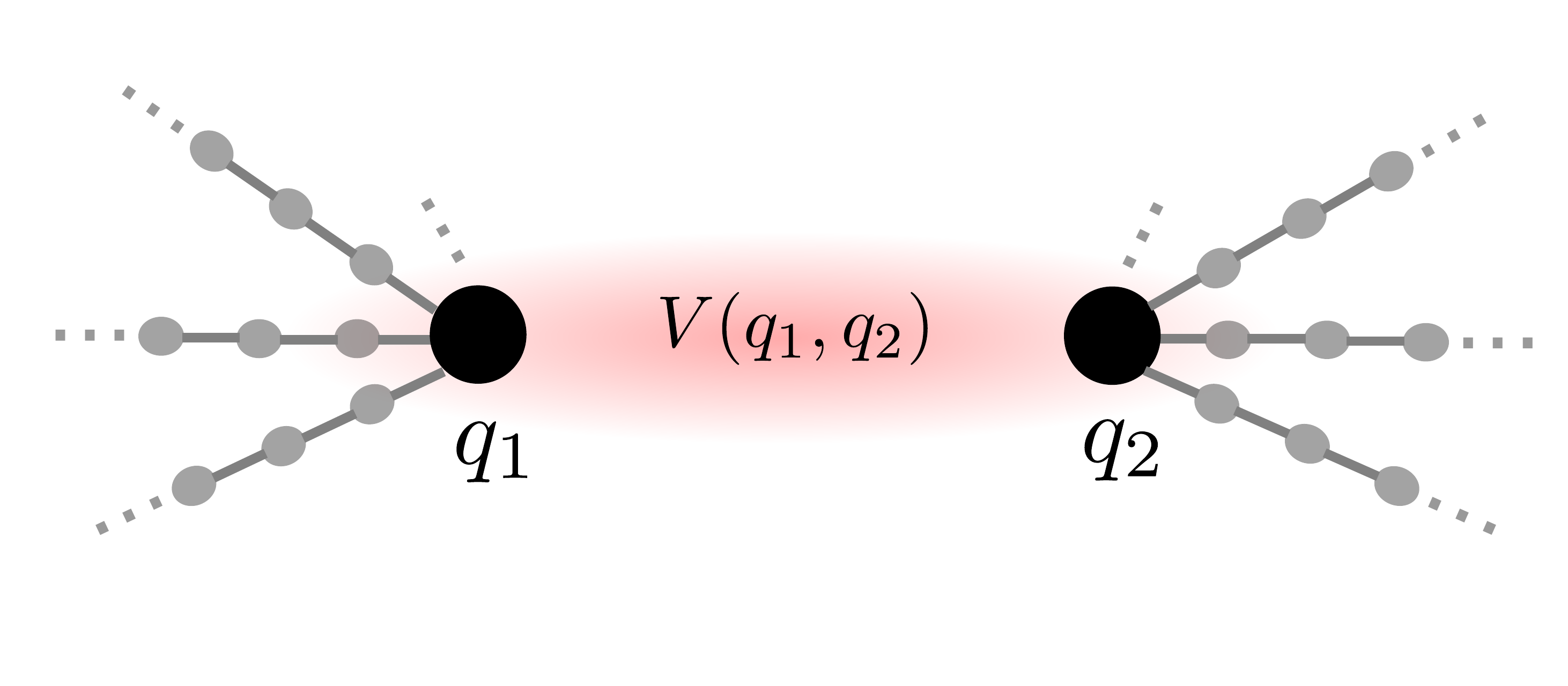}
\caption{Several transmission lines attached to the open subsystem.}
\label{SeveralLines}
\end{figure}
Finally, we point out that the system in Figure~\ref{SeveralLines} is a combination of the Heat Bath (Figure~\ref{heatbath}) and the single transmission line (Figure~\ref{fig2}). } \hfill $\diamond$
\end{remark}

\subsection{Discrete setting} Now we will consider the closed system illustrated in Figure \ref{fig2} in a temporally discrete setting. The corresponding discrete dynamics is encoded in the discrete Lagrangian \eqref{DiscLagN}, where we choose $\gamma=0$, which in this case reads
\[
\begin{split}
&L(\mathfrak{q}_k,\mathfrak{q}_{k+1})= \sum_{a=1}^2 L_d^s(q_a^k, q_a^{k+1}) + \sum_{a=1}^2 \sum_{j=1}^{(\mathcal{N}-2)/2} L_d^{\tiny\mbox{tl}}((Q_a^j)^k, (Q_a^j)^{k+1})+ L_d^{\tiny\mbox{int}}(q_1^{k+1},q_2^{k+1},(Q_1^1)^{k+1},(Q_2^1)^{k+1}),
\end{split}
\]
where $k$ denotes the discrete time index. Similar to the continuous counterpart, $L_d^s$ encodes the dynamics of the open system, $L_d^{\tiny\mbox{tl}}$ encodes the dynamics of the environment and $L_d^{\tiny\mbox{int}}$ encodes the  system-environment interaction:
\begin{subequations}\label{DiscLagLines}
\begin{align}
L_d^s(q_a^k, q_a^{k+1}) &= \frac{1}{2h}m_a (q_a^{k+1}-q_a^k)^2 - hV(q_1^{k+1},q_2^{k+1}), \label{DiscLagLines:a}\\
L_d^{\tiny\mbox{tl}}((Q_a^j)^k, (Q_a^j)^{k+1}) &= \frac{1}{2h}M((Q_a^j)^{k+1}-(Q_a^j)^{k})^2 \label{DiscLagLines:b}\\
&\hspace{3.5cm} - \frac{1}{2}h\lambda_a((Q_a^{j+1})^{k+1}-(Q_a^{j})^{k+1})^2, \nonumber\\
L_d^{\tiny\mbox{int}}(q_1^{k+1},q_2^{k+1},(Q_1^1)^{k+1},(Q_2^1)^{k+1}) &=- \frac{1}{2}h\lambda_1 ((Q_1^1)^{k+1} - q_1^{k+1})^2
-\frac{1}{2}h\lambda_2 ((Q_2^1)^{k+1} - q_2^{k+1})^2.\label{DiscLagLines:c}
\end{align}
\end{subequations}
Furthermore, the discrete Euler-Lagrange equations DELEqs$_{\tiny\mathcal{N}}^0$ \eqref{DiscELEqsN} are:
\begin{subequations}\label{DisWave}
\begin{align}
m_a\frac{q_a^{k+1}-2q_a^{k}+q_a^{k-1}}{h^2}+\der_{q_a} V(q_1^{k},q_2^{k}) - \lambda_a((Q_a^1)^k-q_a^{k}) &=0, \label{DisWave01}\\
M\frac{(Q_a^j)^{k+1}-2(Q_a^j)^{k}+(Q_a^j)^{k-1}}{h^2} - \lambda_a((Q_a^{j+1})^{k} -2(Q_a^{j})^{k}+(Q_a^{j-1})^{k})&=0, \label{DisWave02}
\end{align}
\end{subequations} 
for $k=1,...,N-1$ and $j=2,...,(\mathcal{N}-2)/2-1$. It is easy to check (for details see \cite{DavidsPaper}) that discrete wave-like equation \eqref{DisWave02} can be satisfied with following solution:
\[
(Q^j)^k := q^{\tiny\mbox{in}}(x_j + v t_{k}) + q^{\tiny\mbox{out}}(x_j - v t_{k})
\]
in terms of inward and outward discrete traveling waves where the velocity $v = \frac{h_x}{h}$ is defined in terms of the spatial and temporal slices and $t_k=kh$, $k=0,...,N.$ According to this, it is easy to check that
\[
q^{\tiny\mbox{out}}(x_j - v t_{k+1}) = q^{\tiny\mbox{out}}(x_j - vh - vt_{k}) =  q^{\tiny\mbox{out}}(x_{j-1} - v t_k).
\]
Similar to the continuous analogue, one finds  $v_a = \sqrt{K_a/\rho}$. By making the same crucial assumption of no reflection in the transmission line, i.e. $q^{\tiny\mbox{in}}(x_j - v t_{k}) = 0$, we can see the following identity:
\begin{equation}
(Q_a^{j+1})^k:= q^{\tiny\mbox{out}}(x_{j+1} - v_a t_{k}) = q^{\tiny\mbox{out}}(x_j - v_a t_{k-1}) = (Q_a^j)^{k-1}
\label{SpaTemp2}
\end{equation}
which converts spatial differences into temporal differences. As a result of that, the equation \eqref{DisWave01} can be re-written in the following form:
\begin{equation}
m_a\frac{q_a^{k+1}-2q_a^{k}+q_a^{k-1}}{h^2}+\der_{q_a} V(q_1^{k},q_2^{k}) + D_a\frac{q_a^k-(Q_a^0)^{k-1}}{h} = 0 \label{DisDissLag1}
\end{equation} 
with $D_a = \sqrt{K_a\rho}$ being the damping rate. Since $Q_a^0 \equiv q_a$, we obtain 
\begin{equation}
m_a\frac{q_a^{k+1}-2q_a^{k}+q_a^{k-1}}{h^2}+\der_{q_a} V(q_1^{k},q_2^{k}) + D_a\frac{q_a^k-q_a^{k-1}}{h} = 0 \label{DisDissLag2}.
\end{equation} 
In this example, these are FDELEqs$_n^0$.

The process leading to \eqref{DisDissLag2} corresponds to the Lagrangian Discrete Damping Reduction (DDR) \eqref{DDR:L} for the transmission line system, with $\tilde\Lambda$ determined in \eqref{tildeLambdaLines}. The same discrete forced Euler-Lagrange equations  may be obtained through the discrete Lagrange-d'Alembert principle \eqref{ForcedDEL} by picking $L_d^s(q_a^k,q_a^{k+1})$ \eqref{DiscLagLines:a} and the discrete forces \eqref{DiscExtForces} with $\gamma=0$, as established in diagram \eqref{DiscDiag}.

From \eqref{DiscLagLines}, we apply the discrete Legendre transform \eqref{DLT}, in order to obtain DHEqs$_{\tiny\mathcal{N}}^0$:

\begin{subequations}\label{MMLines}
\begin{align}
&p_a^{(k,-)} =\quad\,\,\,\,\,\, m_a\frac{(q_a^{k+1} - q_a^{k})}{h}, \label{MMLines:a}\\
&p_a^{(k+1,+)}=\quad m_a\frac{q_a^{k+1} - q_a^{k}}{h} -h\der_{q_a} V(q_1^{k+1},q_2^{k+1}) + h\lambda_a((Q_a^1)^{k+1} - q_a^{k+1}),  \label{MMLines:b}\\
&(P_a^j)^{(k,-)}=\,\,\,\,M\frac{(Q_a^j)^{k+1} - (Q_a^j)^{k}}{h},\label{MMLines:c}\\
&(P_a^j)^{(k+1,+)}=M\frac{(Q_a^j)^{k+1} - (Q_a^j)^{k}}{h}+h\lambda_a((Q_a^{j+1})^{k+1}-2(Q_a^{j})^{k+1}+(Q_a^{j-1})^{k+1}). \label{MMLines:d}
\end{align}
\end{subequations}
It is easy to observe that the momentum mathching $(p_a)^{(k,-)} = p_a^{(k,+)}$ and $(P_a^j)^{(k,-)} = (P_a^j)^{(k,+)}$ reproduce the discrete Euler-Lagrange equation \eqref{DisWave01} and \eqref{DisWave02}, respectively. Finally, the equations \eqref{MMLines:a}, \eqref{MMLines:b} can be expressed in the following form:

\[
\begin{split}
q_a^{k+1} &= q_a^k +h\frac{p_a^k}{m_a},\\
p_a^{k+1} &= p_a^k -h\der_{q_a} V(q_1^{k+1},q_2^{k+1}) - h\,D_a\frac{p^k_a}{m_a},
\end{split}
\]
given the appropriate elimination of the discrete dynamics of the enviroment \eqref{MMLines:c}, \eqref{MMLines:d} based  on the conversion of spatial difference to temporal difference encoded in the equation \eqref{SpaTemp2}. This accounts for the Hamiltonian DDR \eqref{DDR:H}, and the previous equations are the DFHEqs$_n^0$ in this example.
 They can be equivalently obtained from forced momentum matching \eqref{DForcedLT}, i.e.: 
\[
\begin{split}
p_a^{(k,-)} &=\,\,\, m_a\frac{(q_a^{k+1} - q_a^{k})}{h}, \\
p_a^{(k+1,+)}&=\,\,\, m_a\frac{q_a^{k+1} - q_a^{k}}{h} -h\der_{q_a} V(q_1^{k+1},q_2^{k+1}) - D_a(q_a^{k+1} - q_a^{k+1}),
\end{split}
\]
as established by diagram \eqref{DiscDiag}. This provides its commutativity in the Rubin model case, as set in Theorem \ref{DiscTheo}.

Moreover, as mentioned in Remark \ref{ForcedSemiImpliRemark}, the obtained integrators are semi-implicit in nature which is a direct result of the DDR of the symplectic integrators for the system-environment as a whole (closed system) given in equations \eqref{DisWave}.

In the following we compare the accuracy of the approximations of different integrators for a binary oscillator in a linear dissipative environment illustrated in Figure {\ref{numericals}(a), which corresponds to the potential $V(q_1,q_2)=\frac{1}{2}k(q_1-q_2)^2$. 
\begin{figure}[!htb]
\includegraphics[scale=0.4]{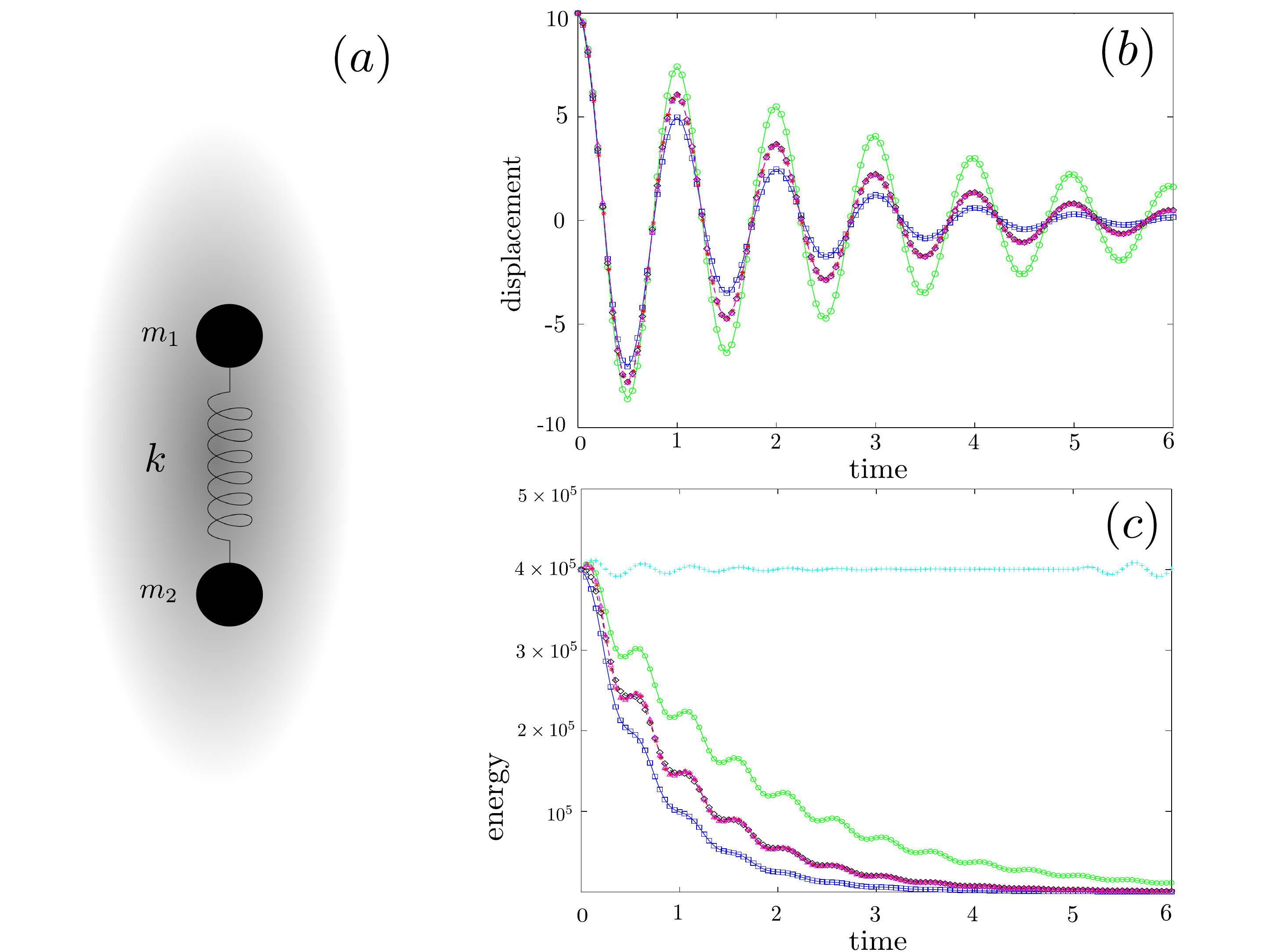}
\caption{$(a)$ - The binary oscillator in a linear dissipative environment. $(b)$  Dynamics of the displacement of the $m_1$ in time with the initial condition $q_1(0) = 10, q_2(0) = -10 $, and the parameters $m_1 = m_2 = 100, k = 10000$ and a characteristic damping rate $D = 100$. The integrators constructed by explicit Euler, implicit Euler and semi-implicit Euler are represented by green circles, blue squares, and red stars, respectively. The numerically exact solution and symplectic integration associated to the closed system of transmission line with the parameters $M =1, \lambda = 2000$ are represented with black diamonds and magenta triangles, both of which almost coincide with semi-implicit approximation. $(c)$ Depicts the energy decays of the binary system in time with the same symbols described in $(b)$. In addition the full energy of the closed system is calculated by the symplectic integrator and represented by cyan plus symbols. As one can see the total energy of the system-environment is a symmetry of the dynamics and is preserved up to bounded oscillations under symplectic integration.}
\label{numericals}
\end{figure}
Figures \ref{numericals} (b) and (c) depict the approximate dynamics of the displacement of $m_1$  and the energy decay of the binary oscillator respectively, with the initial condition $q_1(0) = 10, q_2(0) = -10$. 

\newpage

\section{Conclusions}

We prove the commutativity of the branches {\it Legendre Transform + Damping Reduction} and {\it Damping Reduction +Legendre transform} as shown in  Figure \ref{Sketch1} (left), connecting Lagrangian and Hamiltonian descriptions of open/closed dynamics, for a system of $\mathcal{N}$ particles with quadratic interaction and both continuous and discrete versions of mechanics. We consider the damping as a physical phenomenon due to the interaction between the given open system (linearly damped) and its environment, forming both a closed system. The Damping Reduction is considered as the mapping from the equations of motion of the closed system to those of the open one (i.e. closed $\Flder$ open dynamics).  As a natural instrument for the dynamical description of forced systems (open), we employ the Lagrange-d'Alembert principle. Although this commutativity might seem natural when established at a continuous level, it is not at a discrete level. We prove it for a family of discretisations depending on a parameter $\gamma\in[0,1]$, accounting for the major contribution of the paper.

We observe that the applicability of the Damping Reduction in the exposed terms (i.e.~Figure~\ref{Sketch1} (right)) depends on the distribution of the particles, represented by the matrix $\Lambda$. We provide two particular examples  (heat bath $\Lambda_{\tiny\mbox{hb}}$ and trasmission lines $\Lambda_{\tiny\mbox{tl}}$) where it can be performed. The questions for what kind of $\Lambda$ the reduction can be undertaken, or whether any $\Lambda$ can be transformed to $\Lambda_{\tiny\mbox{hb,tl}}$ via linear change of variables, remain open, and also interesting lines of investigation.

The Damping Reduction at a discrete level yields particular numerical schemes approximating the dynamics of linearly damped systems. These schemes are not symplectic anymore, but preserve some of the features of their symplectic counterparts (for instance the semi-implicitness in some cases) which are indeed obtained in a variational way. We display some simulations for the transmission line example, and observe a better performance over other usual integrators (both dynamically and energy wise), in particular the Euler schemes. This is a new example showing that the integrators with a variational origin perform superiorly when approximating the energy evolution of the systems, even though they are not conservative.

\medskip\medskip

{\bf Acknowledgments}: This work has been funded by the EPSRC project: `'Fractional Variational Integration and Optimal Control''; ref: EP/P020402/1.

\end{document}